\documentclass[twocolumn,amsmath,showpacs,nofootinbib]{revtex4-1}
\usepackage{graphicx}
\usepackage{dcolumn}
\usepackage{bm}
\usepackage{color} 
\usepackage{slashed}
\begin{document}
\newcommand{\hs}{\hspace*{0.5cm}}
\newcommand{\vs}{\vspace*{0.5cm}}
\newcommand{\be}{\begin{equation}}
\newcommand{\ee}{\end{equation}}
\newcommand{\bea}{\begin{eqnarray}}
\newcommand{\eea}{\end{eqnarray}}
\newcommand{\ben}{\begin{enumerate}}
\newcommand{\een}{\end{enumerate}}
\newcommand{\bde}{\begin{widetext}}
\newcommand{\ede}{\end{widetext}}
\newcommand{\nn}{\nonumber}
\newcommand{\crn}{\nonumber \\}
\newcommand{\Tr}{\mathrm{Tr}}
\newcommand{\non}{\nonumber}
\newcommand{\noi}{\noindent}
\newcommand{\al}{\alpha}
\newcommand{\la}{\lambda}
\newcommand{\bet}{\beta}
\newcommand{\ga}{\gamma}
\newcommand{\va}{\varphi}
\newcommand{\om}{\omega}
\newcommand{\pa}{\partial}
\newcommand{\+}{\dagger}
\newcommand{\fr}{\frac}
\newcommand{\bc}{\begin{center}}
\newcommand{\ec}{\end{center}}
\newcommand{\Ga}{\Gamma}
\newcommand{\de}{\delta}
\newcommand{\De}{\Delta}
\newcommand{\ep}{\epsilon}
\newcommand{\varep}{\varepsilon}
\newcommand{\ka}{\kappa}
\newcommand{\La}{\Lambda}
\newcommand{\si}{\sigma}
\newcommand{\Si}{\Sigma}
\newcommand{\ta}{\tau}
\newcommand{\up}{\upsilon}
\newcommand{\Up}{\Upsilon}
\newcommand{\ze}{\zeta}
\newcommand{\ps}{\psi}
\newcommand{\Ps}{\Psi}
\newcommand{\ph}{\phi}
\newcommand{\vph}{\varphi}
\newcommand{\Ph}{\Phi}
\newcommand{\Om}{\Omega}
\newcommand{\AdrHEPC}{Phenikaa Institute for Advanced Study and Faculty of Basic Science, Phenikaa University, Yen Nghia, Ha Dong, Hanoi 100000, Vietnam}

\title{Canonical seesaw implication for two-component dark matter} 

\author{Phung Van Dong} 
\email{Corresponding author.\\ dong.phungvan@phenikaa-uni.edu.vn}
\author{Cao H. Nam} 
\email{nam.caohoang@phenikaa-uni.edu.vn}
\author{Duong Van Loi} 
\email{loi.duongvan@phenikaa-uni.edu.vn}
\affiliation{\AdrHEPC}
 
\date{\today}

\begin{abstract}
We show that the canonical seesaw mechanism implemented by the $U(1)_{B-L}$ gauge symmetry provides two-component dark matter naturally. The seesaw scale that breaks $B-L$ defines a residual gauge symmetry to be $Z_6=Z_2\otimes Z_3$, where $Z_2$ leads to the usual matter parity, while $Z_3$ is newly recognized, transforming quark fields nontrivially. The dark matter components---that transform nontrivially under the matter parity and $Z_3$, respectively---can gain arbitrary masses, despite the fact that the $Z_3$ dark matter may be heavier than the light quarks $u,d$. This dark matter setup can address the XENON1T anomaly recently observed and other observables, given that the dark matter masses are nearly degenerate, heavier than the electron and the $B-L$ gauge boson $Z'$, as well as the fast-moving $Z_3$ dark matter has a large $B-L$ charge, while the $Z'$ is viably below the beam dump experiment sensitive regime.       
\end{abstract}


\maketitle

{\it Motivation.} Neutrino mass and dark matter are the two big questions in science, which require the new physics beyond the standard model \cite{Tanabashi:2018oca}. 

It is well established that the canonical seesaw mechanism can generate appropriate small neutrino masses through the exchange of heavy Majorana right-handed neutrino singlets, $\nu_{aR}$ for $a=1,2,3$, added to the standard model  \cite{Minkowski:1977sc,GellMann:1980vs,Yanagida:1979as,Glashow:1979nm,Mohapatra:1979ia,Mohapatra:1980yp,Lazarides:1980nt,Schechter:1980gr,Schechter:1981cv}. However, the canonical seesaw in its simple form does not naturally address the dark matter issue, unless some dark matter stability condition or parameter finetuning is {\it ad hoc} imposed. 

The simplest gauge completion of the seesaw mechanism with $U(1)_{B-L}$ can provide a natural origin for the existence of the right-handed neutrinos and the right-handed neutrino mass scale \cite{Davidson:1978pm,Marshak:1979fm,Mohapatra:1980qe}. This work shows that such theory manifestly yields a novel consequence of two-component dark matter, properly solving the recent XENON1T excess \cite{Aprile:2020tmw}.

{\it Description of the model.} Indeed, the full gauge symmetry takes the form, 
\be SU(3)_C\otimes SU(2)_L\otimes U(1)_Y\otimes U(1)_{B-L}.\label{fgsadd1}\ee
Here the right-handed neutrino fields $\nu_{aR}$ transforming under the gauge symmetry as 
\be \nu_{aR}\sim (1,1,0,-1)\ee are required in order to cancel the $[\mathrm{Gravity}]^2 U(1)_{B-L}$ and $[U(1)_{B-L}]^3$ anomalies. Additionally, a scalar singlet transforming under the gauge symmetry as
\be \chi\sim (1,1,0,2) \ee must be presented to break $U(1)_{B-L}$ for the model consistency, simultaneously generating the right-handed neutrino masses or the seesaw scale. 

As usual, let us assign the standard model lepton, quark, and Higgs representations with respect to the new gauge symmetry to be,
\bea && l_{aL}=
\left(\begin{array}{c}
\nu_{aL}\\
e_{aL}
\end{array}\right)\sim \left(1,2,-\fr 1 2,-1\right),\\ 
&& e_{aR}\sim (1,1,-1,-1),\\
&& q_{aL}=
\left(\begin{array}{c}
u_{aL}\\
d_{aL}
\end{array}\right)\sim \left(3,2,\fr 1 6,\fr 1 3 \right),\\
&& u_{aR}\sim \left(3,1,\fr 2 3, \fr 1 3\right),\\
&& d_{aR}\sim \left(3,1,-\fr 1 3, \fr 1 3\right),\\
&& \phi = \left(\begin{array}{c}
\phi^+\\
\phi^0
\end{array}\right)\sim \left(1,2,\fr 1 2,0\right).\eea

The scalar multiplets develop vacuum expectation values (VEVs), such as 
\be \langle \chi\rangle=\fr{\La}{\sqrt{2}},\hs \langle \phi\rangle = \left(\begin{array}{c}
0\\
\fr{v}{\sqrt{2}}
\end{array}\right),\ee satisfying \be \La\gg v=246\ \mathrm{GeV}.\ee 

The Yukawa Lagrangian includes \bea \mathcal{L} &\supset& h^{\nu}_{ab}\bar{l}_{aL}\tilde{\phi}\nu_{bR}+\fr{1}{2}f^{\nu}_{ab}\bar{\nu}^c_{aR}\chi \nu_{bR}+H.c.\crn
&\supset& -\fr 1 2 (\bar{\nu}_{aL}\ \bar{\nu}^c_{aR})
\left(\begin{array}{cc}
0 & m_{ab} \\
m_{ba} & M_{ab}\end{array}\right)
\left(\begin{array}{c}
\nu^c_{bL}\\
\nu_{bR}\end{array}\right)+H.c., \eea where \be m_{ab}=-h^\nu_{ab} \fr{v}{\sqrt{2}},\hs M_{ab}=-f^\nu_{ab} \fr{\La}{\sqrt{2}}.\ee

Hence, the canonical seesaw is naturally recognized in the $U(1)_{B-L}$ gauge completion given that $v\ll \La$ or $m\ll M$, yielding the observed neutrino $(\sim\nu_{aL})$ masses to be
\be m_\nu =-m M^{-1}m^T=h^\nu (f^\nu)^{-1}(h^\nu)^T\fr{v^2}{\sqrt{2}\La},\ee while the heavy neutrinos ($\sim \nu_{aR}$) obtain large masses at the $B-L$ breaking scale, $M\sim \La$. 

The neutrino oscillation data implies $m_{\nu}\sim 0.1$ eV \cite{Tanabashi:2018oca}, which leads to \be \La\sim [(h^\nu)^2/f^\nu] 10^{14}\ \mathrm{GeV}.\ee The seesaw scale $\La$ is close to the grand unification scale if $(h^\nu)^2/f^\nu\sim 1$. If $(h^\nu)^2/f^\nu$ is sufficiently small, say $f^\nu\sim1$ and $h^\nu\sim 10^{-5.5}$--$10^{-5}$ proportional to the electron Yukawa coupling, we derive $\La\sim 1$--10 TeV, in agreement to the collider bounds \cite{Tanabashi:2018oca}. 

All the above results have been established in the literature. However, a proper realization of residual gauge symmetry of $B-L$ and its implication for dark matter have not emerged yet. Let us call the reader's attention to previous works \cite{Krauss:1988zc,Martin:1992mq,Batell:2010bp, Ma:2015xla,Ma:2015mjd,Hirsch:2017col,Bonilla:2018ynb,Heeck:2013rpa,Cai:2018nob,Nanda:2019nqy} relevant to this proposal.

{\it Residual symmetry and dark matter.} The symmetry breaking scheme is obtained as 
\bc
\begin{tabular}{c}
$SU(3)_C\otimes SU(2)_L\otimes U(1)_Y\otimes U(1)_{B-L}$\\
$\downarrow \La $\\
$SU(3)_C\otimes SU(2)_L\otimes U(1)_Y\otimes R$\\
$\downarrow v$\\
$SU(3)_C\otimes U(1)_Q\otimes R$
\end{tabular}
\ec Here the electric charge is related to the isospin and hypercharge as $Q=T_3+Y$. $R$ is a residual symmetry of $U(1)_{B-L}$ that conserves the $\chi$ vacuum, although this vacuum $\langle \chi\rangle =\La/\sqrt{2}\neq 0$ breaks $B-L$ by two unit. As being a $U(1)_{B-L}$ transformation, $R=e^{i\al (B-L)}$ where $\al$ is a transforming parameter. The vacuum conservation condition $R\langle \chi\rangle = \langle \chi\rangle $ leads to $e^{i\al (2)}=1$, or equivalently $\al=k \pi$ for $k$ integer. Hence, the residual symmetry is 
\be R=e^{i k\pi (B-L)}=(e^{i\pi (B-L)})^k.\ee It is noted that the transformation with $k$ is conjugated to that with $-k$, i.e. $R^\dagger = (e^{i\pi(B-L)})^{-k}=R^{-1}$. 

\begin{table}[h]
\bc
\begin{tabular}{lccc}
\hline\hline
Field & $(\nu,e)$ & $(u,d)$ & $(\phi,\chi,A)$\\
\hline
$R$ & $(-1)^{-k}$ & $e^{ik\pi/3}$ & 1
\\
\hline\hline
\end{tabular}
\caption[]{\label{tab1} $R$ values of leptons, quarks, and bosons, where the generation and left/right chirality indices are omitted since the relevant fields have the same $R$ value.}
\ec
\end{table}   
Let us commonly denote $A$ to be the gauge fields associated with the gauge symmetry in (\ref{fgsadd1}). The $R$ values of all fields are collected in Table~\ref{tab1}.  From this table, we derive that $R=1$ for the minimal value of $|k|=6$ and for every field, except the identity $k=0$. Hence, the residual symmetry $R$ is automorphic to \be Z_6=\{1,p,p^2,p^3,p^4,p^5\},\ee where $p\equiv e^{i\pi (B-L)}$ and $p^6=1$. Further, we factorize \be Z_6=Z_2\otimes Z_3,\ee where 
\be Z_2=\{1,p^3\}\label{bd2}\ee
is the invariant (or normal) subgroup of $Z_6$, while 
\be Z_3=Z_6/Z_2=\{Z_2, \{p,p^4\},\{p^2,p^5\}\} \label{qg}\ee
is the quotient group of $Z_6$ by $Z_2$. Thus, the theory automatically conserves both residual symmetries $Z_2$ and $Z_3$ after symmetry breaking.   

\begin{table}[h]
\bc
\begin{tabular}{lccc}
\hline\hline
Field & $(\nu,e)$ & $(u,d)$ & $(\phi,\chi,A)$\\
\hline
$1$ & 1 & 1 & 1 \\
$p^3$ & $-1$ & $-1$ & 1 \\
\hline
$Z_2$ & $\underline{1}'_{2}$ & $\underline{1}'_{2}$ & $\underline{1}_{2}$\\
\hline
$p$ & $-1$ & $-w^2$ & 1\\
$p^4$ & 1 & $w^2$ & 1\\
\hline
$p^2$ & 1 & $w$ & 1\\
$p^5$ & $-1$ & $-w$ & 1\\
\hline
$Z_3$ & $\underline{1}_{3}$ & $\underline{1}'_{3}$ & $\underline{1}_{3}$ \\ 
\hline\hline
\end{tabular}
\caption[]{\label{tab2} Field representations under the residual symmetry $R=Z_2\otimes Z_3$.}
\ec
\end{table}   
We affix the subscripts $_{2,3}$ to a $N$-dimensional representation $\underline{N}$ if it is viable, say $\underline{N}_{2,3}$, in order to indicate to those of $Z_{2,3}$, respectively. The field representations under $Z_2$ and $Z_3$ are computed in Table \ref{tab2}, where $w\equiv e^{i2\pi/3}$ is the cube root of unity. Here note that $Z_2$ has two (1-dimensional) irreducible representations, $\underline{1}_{2}$ according to $p^3=1$ and $\underline{1}'_{2}$ according to $p^3=-1$, whereas $Z_3$ has three (1-dimensional) irreducible representations, $\underline{1}_{3}$ according to $(p^2,p^5)=(1, 1)$ or $(1,- 1)$, $\underline{1}'_{3}$ according to $(p^2,p^5)=(w,w)$ or $(w,-w)$, and $\underline{1}''_{3}$ according to $(p^2,p^5)=(w^2,w^2)$ or $(w^2,-w^2)$, which are independent of $p^3$ values, $1$ or $-1$, that identify $Z_6$ elements in a coset of the quotient group, respectively.\footnote{The nontrivial representations of $Z_3$ obey
$\underline{1}'_{3}\otimes \underline{1}''_{3}=\underline{1}_{3},\ (\underline{1}'_{3})^3=(\underline{1}''_{3})^3=\underline{1}_{3},\ (\underline{1}'_{3})^*=\underline{1}''_{3}=(\underline{1}'_{3})^2$, and $(\underline{1}''_{3})^*=\underline{1}'_{3}=(\underline{1}''_{3})^2$, whereas that of $Z_2$ satisfies $(\underline{1}'_{2})^2=\underline{1}_{2}$ and $(\underline{1}'_{2})^*=\underline{1}'_{2}$.} The representation $\underline{1}''_{3}$ is not presented for the existing fields, but the antiquarks $(u^c,d^c)$ belong to $\underline{1}''_{3}$ under $Z_3$. 

For brevity, the quotient group can be defined as 
\be Z_3=\{[1],[p^2],[p^4]\},\ee 
where each (coset) element $[x]$ consists of two elements of $Z_6$, the characteristic $x$ and the other $p^3 x$, as multiplied by $p^3$. Hence, $[1]=[p^3]=Z_2$, $[p]=[p^4]=\{p,p^4\}$, and $[p^2]=[p^5]=\{p^2,p^5\}$. Because of $[p^4]=[p^2]^2=[p^2]^*$, the $Z_3$ group is completely generated by \be [p^2]=[e^{i2\pi(B-L)}]=[w^{3(B-L)}].\ee That said, the $Z_3$ irreducible representations $\underline{1}_{3}$, $\underline{1}'_{3}$, and $\underline{1}''_{3}$ are simply determined by $[p^2]=[1]\rightarrow 1$, $[p^2]=[w]\rightarrow w$, and $[p^2]=[w^2]\rightarrow w^2$, respectively. Here, the intermediate $Z_6$ representations $[r]$ consists of $r$ and $\pm r$ as multiplied by $p^3=\pm 1$ respectively, which are homomorphic to that of $Z_3$, $[r]=\{r,\pm r\}\rightarrow r$.   

Since the spin parity $P_S=(-1)^{2s}$ is always conserved by the Lorentz symmetry, we can conveniently multiply the residual symmetry $R=Z_2\otimes Z_3$ with spin-parity group $S=\{1,P_S\}$ to perform 
\be R\rightarrow R\otimes S=(Z_2\otimes S)\otimes Z_3,\ee where $Z_3$ is retained as the quotient group. The new invariant subgroup $Z_2\otimes S$ defines a matter parity 
\be P_M=p^3\times P_S = (-1)^{3(B-L)+2s},\ee analogous to the $R$-parity in supersymmetry. 

Because of $P^2_M=1$, we have $P=\{1,P_M\}$ to be a group of matter-parity symmetry by itself, which is an invariant subgroup of $Z_2\otimes S$. Therefore, we factorize 
\be R\otimes S= [(Z_2\otimes S)/P]\otimes P\otimes Z_3.\ee Here $(Z_2\otimes S)/P=\{P,\{p^3,P_S\}\}$ is conserved, if $P_M$ is conserved. Therefore, instead of $R\otimes S$, we can consider an alternative residual symmetry which is contained in 
\be R\otimes S \supset P\otimes Z_3, \ee where the quotient group $(Z_2\otimes S)/P$ is neglected, since the theory automatically preserves it. Of course, the theory conserves both $P$ and $Z_3$, under which the representations under these groups are given in Table \ref{tab3}, in which the subscript $_P$ indicates to the representations of the matter parity group ($P$).    

\begin{table}[h]
\bc
\begin{tabular}{lccc}
\hline\hline
Field & $(\nu,e)$ & $(u,d)$ & $(\phi,\chi,A)$\\
\hline
$P_M$ & 1 & 1 & 1 \\
$P$ & $\underline{1}_{P}$ & $\underline{1}_P$ & $\underline{1}_P$\\
\hline
$[p^2]$ & $1$ & $w$ & $1$\\
$Z_3$ & $\underline{1}_{3}$ & $\underline{1}'_{3}$ & $\underline{1}_{3}$ \\ 
\hline\hline
\end{tabular}
\caption[]{\label{tab3} Field representations under the alternative residual symmetry $P\otimes Z_3$, where note that the matter parity group $P$ is isomorphic to a $Z_2$ group generated by $P_M$, while the quotient group $Z_3$ is generated by $[p^2]$.}
\ec
\end{table}   

Hence, the model provides a natural stability mechanism for two-component dark matter, in which a dark matter component transforms nontrivially under the matter parity group $P \cong Z_2$ [which should not be confused with the first $Z_2$ in (\ref{bd2})], i.e. in $\underline{1}'_P$ of $P$ characterized by $P_M=-1$, while the remaining dark matter component transforms nontrivially under the quotient group $Z_3$, i.e. in $\underline{1}'_3$ or $\underline{1}''_3$ of $Z_3$ characterized by $[p^2]=[w]\rightarrow w$ or $[p^2]=[w^2]\rightarrow w^2$, respectively. 

First, to ensure the residual symmetry $Z_6$, i.e. $p^6=\exp[i6\pi (B-L)]=1$, every dark field should possess a $B-L$ charge, such that $3(B-L)$ is integer. Thus, $[p^2]$ is nontrivial if and only if \[B-L=\pm \fr 1 3 + n_1=\pm \fr 1 3,\pm \fr 2 3,\pm \fr 4 3,\pm \fr 5 3,\cdots\] for a generic field. Moreover, $P_M$ is odd if and only if \[B-L=\fr{1+2n_2}{3}=\pm \fr 1 3,\pm 1,\pm \fr 5 3,\pm \fr 7 3,\cdots\] for bosonic field, and \[B-L=\fr{2 n_3}{3}=0,\pm \fr 2 3,\pm \fr 4 3,\pm 2,\cdots\] for fermionic field. The charge parameters $n_{1,2,3}$ are arbitrarily integer. Thus, $B-L$ is quantized with basic periods $1$ and $2/3$ resulting from cyclic property of the residual symmetries $Z_3$ and $P$, respectively. Additionally, the opposite signs indicate that a field and its conjugation belong to the same kind of dark matter, i.e., nontrivial under $Z_3$, $P$, or both. 

To find the minimal realistic setup of two-component dark matter, we demand that the dark sector contains spin-0 bosonic and/or spin-$1/2$ fermionic fields, transforming as standard model singlets. Additionally, the three kinds of dark fields, $P$, $Z_3$, and both, each have a field, called $F_1$, $F_2$, and $\Phi$, respectively, where $F_1\sim (-1,1)$, $F_2\sim (1,w/w^2)$, and $\Phi\sim (-1,w/w^2)$ given in terms of $(P_M,[p^2])$ nontrivially transform under $P$, $Z_3$, and both, respectively. 
\begin{table}[h]
\bc
\begin{tabular}{lcc}
\hline\hline
Field & Scalar & Fermion \\
\hline
$[B-L](F_1)$ & $1+2n_1$ & $2n_1$ \\
$[B-L](F_2)$ & $\pm \fr 2 3 + 2 n_2$ & $\pm \fr 1 3 + 2n_2$ \\
$[B-L](\Phi)$ & $\pm \fr 1 3 +2 n_3$ & $\pm \fr 2 3 + 2 n_3$ \\
\hline\hline
\end{tabular}
\ec
\caption[]{\label{ntdt101p} $B-L$ charge of the dark field kinds that depends on which statistics they obey.}
\end{table}
Hence their $B-L$ charge is determined in Table \ref{ntdt101p} dependent on which type, either a scalar or a fermion, they are.\footnote{There possibly exist renormalizable couplings of $\chi$ to dark fields, such as $\chi F^2_1$, $\chi F^3_2$, $\chi F_1 F_2 \Phi$, and $\chi F_2 \Phi^2$, with appropriate $B-L$ charges, besides the usual ones $(\chi^*\chi) (X^* X)$ for $X=F_1, F_2,\Phi$, given that all the dark fields are scalar.} Obviously, such a dark field can possess a minimal $B-L$ charge, while its real $B-L$ charge is obtained by the cyclic property. Therefore, $F_1$ has either $B-L=0$ as a fermion or $\pm1$ as a scalar, $F_2$ has either $B-L=\pm1/3$ as a fermion or $\pm2/3$ as a scalar, and $\Phi$ has either $B-L=\pm1/3$ as a scalar or $\pm2/3$ as a fermion. Among these solutions based upon $P\otimes Z_3$, let us assume the simplest (i.e. most minimal $B-L$) dark matter candidates, as summarized in Table \ref{tab4}. Note that the $B-L$ charge of each dark field can deviate from the supplied value by an arbitrary even number---the common multiple of the basic $1$ and $2/3$ periods---that necessarily does not change the $P,Z_3$ representations of the field.\footnote{The actual $B-L$ charges of the dark fields would be phenomenologically determined.} Here, $F$ and $\Phi$ mean fermion and scalar dark fields, respectively. Further, we assume the net mass of $F_1$ and $F_2$ is smaller than that of $\Phi$. In this setup, there is no coupling of a $\chi$ to two dark fields, because of the Lorentz and $U(1)_{B-L}$ invariance. 

\begin{table}[h]
\bc
\begin{tabular}{lcccc}
\hline\hline
Field & $P_M$ & $P$ & $[p^2]$ & $Z_3$ \\
\hline
$F_1\sim (1,1,0,0)$ & $-1$ & $\underline{1}'_P$ & $1$ & $\underline{1}_3$\\
$F_2\sim (1,1,0,1/3)$ & $1$ & $\underline{1}_P$ & $w$ & $\underline{1}'_3$\\
$\Phi\sim (1,1,0,-1/3)$ & $-1$ & $\underline{1}'_P$ & $w^2$ & $\underline{1}''_3$\\ 
\hline\hline
\end{tabular}
\caption[]{\label{tab4} Simplest dark matter candidates implied by the residual symmetry $P\otimes Z_3$, where the $P$ and $Z_3$ representations are determined by the matter parity $P_M$ and the quotient generator $[p^2]$, respectively.}
\ec
\end{table}   

The dark matter component stabilized by $P$ (i.e., $F_1$) can have an arbitrary mass. This is also valid for the dark matter component stabilized by $Z_3$ (i.e., $F_2$), despite the fact that this $F_2$ component may be heavier than the light quarks $u,d$. Notice that, since all $F_2$, $u$, and $d$ transform nontrivially under $Z_3$, this symmetry by itself does not prevent $F_2$ from decay to quarks. However, when $Z_3$ is combined with $SU(3)_C$, the $F_2$ stability is ensured. Prove: Since the $Z_3$ dark matter component is color neutral, it cannot decay to any colored final state, such as those that include single quarks, because of the color conservation. The $SU(3)_C$ conservation requires a color-neutral final state if it results from a dark matter decay, by assumption. Obviously, this color-neutral final state if constructed from quarks must include only combinations of $q^c q$ and/or $qqq$ (and/or their conjugation). It follows that the final state is invariant (i.e. singlet) under $Z_3$ too. Because of the $Z_3$ conservation, such a final state cannot be the product of any $Z_3$ dark matter decay, which leads to a contradiction. In other words, the $SU(3)_C$ and $Z_3$ symmetries jointly suppress the decay of $Z_3$ dark matter component (i.e. stabilized), even if this component has a mass larger than that of quark.   

With this proposal, we have the novel, simplest model for two-component dark matter based upon $F_1$ and $F_2$ self-interacting through a heavier dark field $\Phi$, which is of course implied by the residual symmetry $P\otimes Z_3$, thus the canonical seesaw. [We can have other scenarios for two-component dark matter, if either the alternative solutions of $F_{1,2},\Phi$ or more dark fields are introduced and that a coupling of $\chi$ to two dark fields may arise. But they are complicated and suppressed.] Notice that since $F_{1,2}$ and $\Phi$ are the standard model singlets, the $U(1)_{B-L}$ dynamics is crucially/sufficiently governing the dark matter observables, besides the known consequences of neutrino mass and baryon asymmetry \cite{Fukugita:1986hr}. 

{\it Seesaw implication for the XENON1T excess.} The XENON1T experiment has recently reported an excess in electronic recoil energy ranging from 1 to 7 keV, peaked around 2.4 keV, having a local statistical significance above 3$\sigma$ \cite{Aprile:2020tmw}. Such signal of electron recoils seems to reveal the existence of a structured dark sector \cite{Takahashi:2020bpq,Kannike:2020agf,Choi:2020udy,Buch:2020mrg,Chen:2020gcl,Bell:2020bes,Du:2020ybt,Su:2020zny,Harigaya:2020ckz,Fornal:2020npv,Alonso-Alvarez:2020cdv,Jho:2020sku,Baryakhtar:2020rwy,Bloch:2020uzh,
Paz:2020pbc,Cao:2020bwd,Lee:2020wmh,Nakayama:2020ikz,Primulando:2020rdk,Lee:2020wmh,Gelmini:2020xir,Jho:2020sku,Bramante:2020zos,Zu:2020idx,Baek:2020owl,Alhazmi:2020fju,Chao:2020yro,DelleRose:2020pbh,Ko:2020gdg,An:2020tcg,Okada:2020evk,Choudhury:2020xui,Arcadi:2020zni,He:2020wjs,Dey:2020sai,Smirnov:2020zwf,Choi:2020kch}.\footnote{For other interpretations, see \cite{Miranda:2020kwy,Lindner:2020kko,AristizabalSierra:2020edu,Boehm:2020ltd,Khan:2020vaf,Bally:2020yid}, for instance.}
Indeed, the authors in \cite{Kannike:2020agf} first showed that in order to fit the excess, the dark matter component that scatters off electrons should be fast moving $v_2\sim 0.03$--$0.25$ for the dark matter mass $m_2\sim 0.1$ MeV to 10 GeV, which exceeds the velocity of cold dark matter $v_1\sim 10^{-3}$. 

This fast dark matter component ($F_2$) may be generated locally as a boosted dark matter from the annihilation or semi-annihilation of the heavier dark matter component ($F_1$), which is nicely implicated by our model. As a matter of fact, the heavier component with the quantum numbers $F_1\sim (1,1,0,0)$ interacting with normal matter only via gravity would dominate the cold dark matter, set by its annihilation or co-annihilation to the lighter dark matter component $F_2$ through the dark matter self-interaction, $\bar{F}_1 F_2\Phi$ plus its conjugation. The lighter component $F_2$ sub-dominates the dark matter abundance, since it strongly couples to the $Z'$ portal which is totally annihilated before freezeout.     

The relevant Lagrangian terms are
\bea \mathcal{L}&\supset&  \bar{F}_1(i\ga^\mu \pa_\mu -m_1) F_1+\bar{F}_2 (i\ga^\mu D_\mu-m_2)F_2\crn
&&+[(D^\mu \Phi)^\dagger (D_\mu \Phi)-m^2_0\Phi^\dagger \Phi]+\bar{e}(i\ga^\mu D_\mu-m_e)e\crn
&&+(h \bar{F}_1 F_2\Phi +H.c.),\eea where $D_\mu=\pa_\mu + i g_{B-L} (B-L) Z'_\mu$ and the dark matter masses obey $m_0>m_1+m_2$ and $m_1>m_2$. Since the $B-L$ charge of $F_1$ is fixed as $B-L=0$, the remaining dark fields can possess more general $B-L$ charges, \be F_2\sim (1,1,0, 1/3+2n),\hs \Phi\sim (1,1,0,-1/3 -2n),\ee for $n=0,\pm1,\pm2,\cdots$, as mentioned.

\begin{figure}[h]
\includegraphics[scale=0.6]{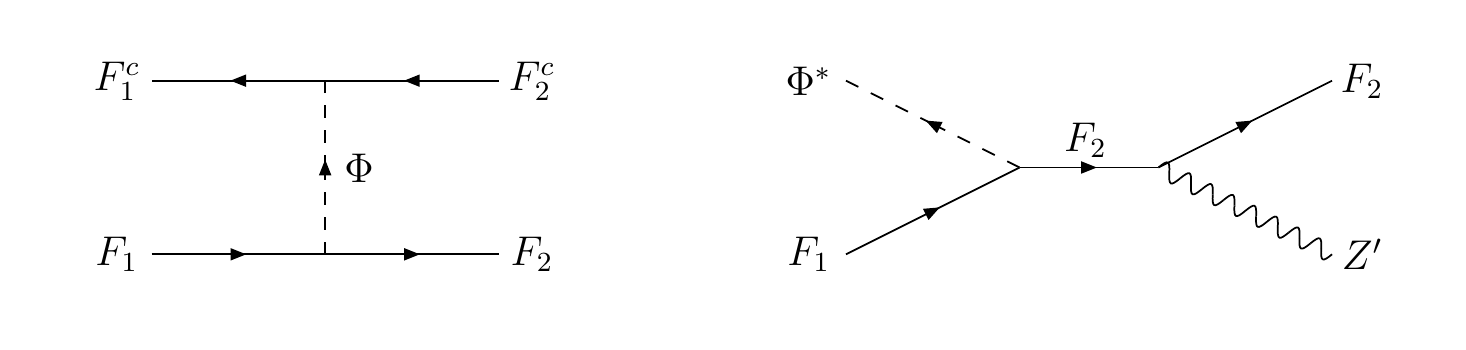}
\caption[]{\label{fig1}Annihilation (left) and co-annihilation (right) processes of $F_1$ that set the cold dark matter density.}
\end{figure}
The relic density of $F_1$ is governed by Feynman diagrams in Fig. \ref{fig1}. The co-annihilation process is only enhanced when the masses of $F_1$ and $\Phi$ are highly degenerate. However, this work signifies $m_0>m_1+m_2\sim 2m_1$ such that the co-annihilation contribution is negligible, where notice that $m_2 \sim m_1$ is used, since $F_2$ is mildly boosted responsible for the XENON1T excess. Hence, the dark matter abundance is given by the $F_1$ annihilation in the left diagram.    

Applying the Feynman rules, we obtain the thermal average cross-section times relative velocity as 
\be
\langle\sigma v_{\mathrm{rel}}\rangle\simeq \frac{|h|^4 (m_1+m_2)^2}{8\pi m_0^4}\left(1-\fr{m^2_2}{m^2_1}\right)^{1/2},
\ee
which relates to the $F_1$ abundance, $\Omega h^2\simeq 0.1 \ \mathrm{pb}/\langle\sigma v_{\mathrm{rel}}\rangle$, where $h$ is the reduced Hubble parameter without confusion. Using the experimental data $\Omega h^2 \simeq 0.12$ \cite{Tanabashi:2018oca} and the Lorentz boost by the left diagram $\ga_2=m_1/m_2=(1-v^2_2)^{-1/2}\approx 1$, i.e. $m_1\approx m_2$, we get the constraint of the dark matter self-coupling to be \be |h| \simeq  7\times 10^{-4}\left(\fr{m_0}{2m_1}\right) \left(\fr{m_1}{1\ \mathrm{MeV}}\right)^{1/2}\left(\fr{\delta m^2}{m^2_1}\right)^{-1/8},\ee which translates to $|h|\gtrsim 10^{-3}$, because of $m_0\gtrsim 2 m_1$, $m_1\gtrsim 1$ MeV, and $\delta m^2/m^2_1= (m^2_1-m^2_2)/m^2_1\simeq v^2_2\sim 1\%$ as shown below. Since the $F_1$ dark matter is thermally generated, its mass has been assumed to be larger than the BBN and CMB bounds $\sim 1$ MeV which are hereafter taken as the electron mass, i.e. $m_1> m_e$ \cite{Sabti:2019mhn}. 

\begin{figure}[h]
\includegraphics[scale=0.8]{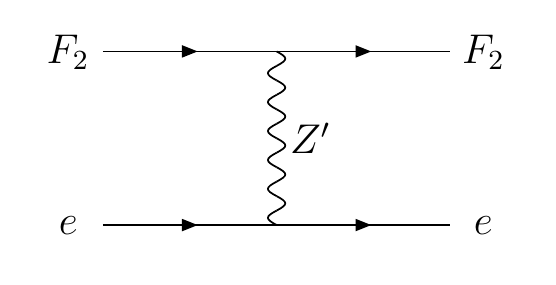}
\caption[]{\label{fig2} Scattering process of the boosted dark matter $F_2$ with electrons in the XENON1T experiment.}
\end{figure}
Of course, at present, $F_2$ is locally generated by the left diagram in Fig. \ref{fig1} which subsequently scatters off electrons in the XENON1T experiment through the diagram in Fig. \ref{fig2}. The recoil energy, deposited by the $F_2$-electron scattering, required to fit the excess is few keV, i.e. $E_R\sim 2.4$ keV. In this model, since the $F_2$ mass satisfies $m_2\approx m_1>m_e$, the electronic recoil energy is estimated as $E_{R}=E_{e'}-E_{e}=2\mu v_{\mathrm{rel}}v_{\mathrm{cm}} < 2m_e v^2_{2}$, where the incoming velocities of $F_2$ and $e$, called $v_2$ and $v_e$ respectively, are assumed to be parallel, $\mu=m_2 m_e/(m_2+m_e)$ is the reduced mass, $v_{\mathrm{rel}}=v_2-v_e$ is the relative velocity, and $v_{\mathrm{cm}}=(m_2 v_2+m_e v_e)/(m_2+m_e)$ is the center-of-mass velocity. Thus, the recoil energy is comparable to the observed value if $F_2$ is boosted with a velocity $v_2\sim 0.1$, in agreement to the best fit in \cite{Kannike:2020agf}. This yields a transferred momentum $q=-E_R/v_{\mathrm{cm}}$, such that $|q| \sim 40\ \mathrm{keV}$. In the limit of mediator mass $m_{Z'}$ to be much larger than the momentum transfer, $m_{Z'}\gg |q|$, the $F_2$-electron scattering cross-section can be written as \cite{Joglekar:2019vzy} 
\be
\sigma_e = \frac{g_{B-L}^4 (1/3+2n)^2 m^2_e}{\pi m^4_{Z'}}.
\ee
This leads to the number of the signal events as related to the scattering cross-section by \cite{Fornal:2020npv,Agashe:2014yua}
\be
\frac{N_{\mathrm{sig}}}{100} = \frac{16 \sigma_e }{3\ \mathrm{pb}}\left(\frac{1\ \mathrm{MeV}}{m_1}\right)^2.\label{nes2d}
\ee    

The mass of $Z'$ boson is $m_{Z'}=2g_{B-L}\La$, where the contribution of the kinetic mixing effect between the $U(1)_Y$ and $U(1)_{B-L}$ gauge bosons is radically small and neglected \cite{Tanabashi:2018oca}. Therefore, the scattering cross-section becomes $\sigma_e = (1/3+2n)^2 m^2_e/(16\pi\La^4)$, which does not explicitly depend on the $Z'$ mass, $m_{Z'}$, and the $Z'$ coupling, $g_{B-L}$, but on the $B-L$ breaking scale $\La= m_{Z'}/2g_{B-L}$, governed by the ratio of the $Z'$ mass to the $Z'$ coupling. Additionally, the new observation is that the cross-section is significantly enhanced by the $B-L$ charge of $F_2$ dark matter as multiplied by $n^2$ as this charge is large, but it is suppressed by the new physics scale to be $1/\La^{4}$. Correspondingly, the event number is proportional to $N_{\mathrm{sig}}/100\sim (10^{-5}n)^2(m_e/m_1)^2(\mathrm{TeV}/\La)^4$, as enhanced by $n^2$, but suppressed by $1/m^2_1\times 1/\La^4$. 

\begin{figure}[h]
\includegraphics[scale=0.33]{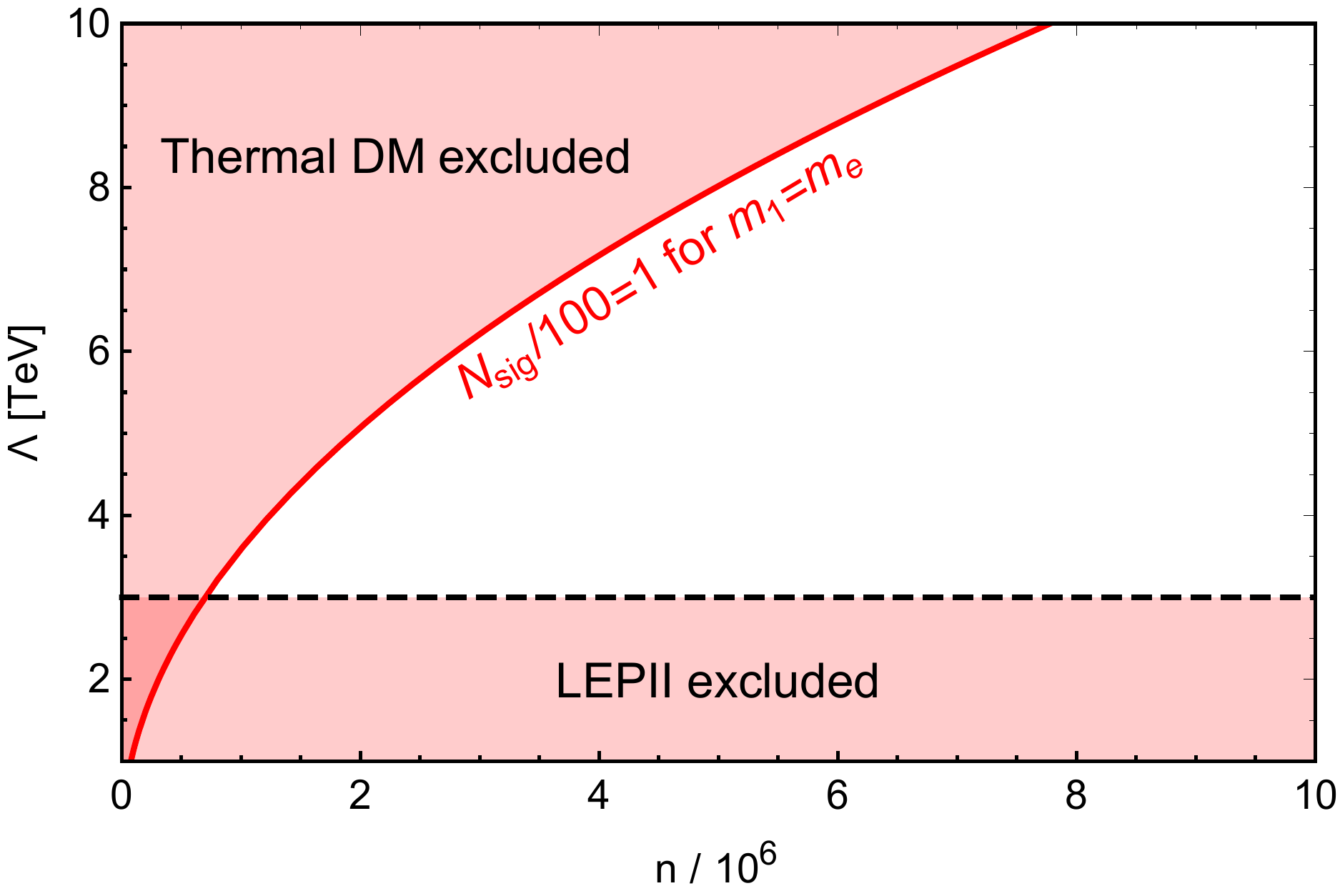}\\ 
\vspace{0.2cm}
\includegraphics[scale=0.33]{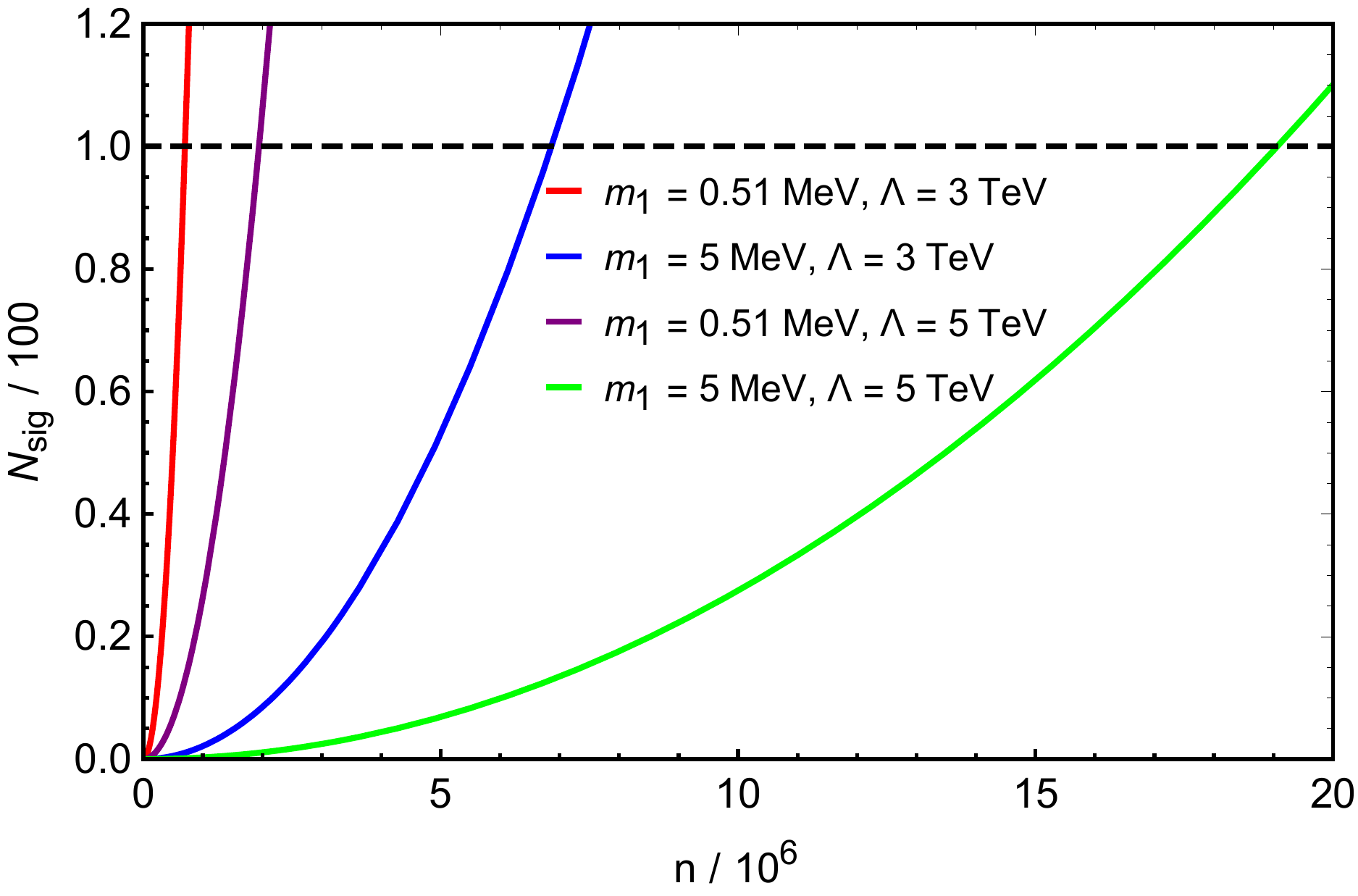}
\caption[]{\label{fig3} Upper panel: Allowed (white) region of $(n,\La)$ determined by (i) contour of $N_{\mathrm{sig}}/100=1$ as a function of $(n,\La)$ according to the WIMP mass limit $m_1=m_e$ and (ii) inclusion of the collider constraint on $\La$. Lower panel: Number of the signal events, $N_{\mathrm{sig}}/100$, significantly enhanced by the fast dark matter charge for the reference values of the thermal dark matter mass and the new physics scale.} 
\end{figure}

The viable regime of the dark matter charge and the new physics scale is determined by making a contour of $N_{\mathrm{sig}}/100=1$ from (\ref{nes2d}) as implied by the XENON1T experiment to be a function of $(n,\La)$ as in Fig.~\ref{fig3} upper panel, taking the lower mass limit for thermal dark matter $m_1=m_e\simeq 0.51$~MeV, where the collider bound $\La=3$~TeV is shown (cf. Appendix~\ref{appdat}), as well as noting that the thermal dark matter excluded region corresponds to $m_1<m_e$. On the other hand, the enhancement of the event number in terms of the dark matter charge is illustrated in Fig.~\ref{fig3} lower panel, taking into account some allowed values of the dark matter mass and the new physics scale from the above regime: \be (m_1,\La)= (0.51,3),\ (0.51,5),\ (5, 3),\ (5, 5),\label{etad1}\ee in $(\mathrm{MeV},\mathrm{TeV})$, respectively. 

Remarks are in order 
\ben
\item
From the figure, the dark matter charge has a lower limit to be large, required in order to explain the number of the signal events, $N_{\mathrm{sig}}/100=1$, i.e. \be |1+6n|\simeq 915.15 \fr{m_1 \Lambda^2}{\text{GeV}^3},\label{dtlt3r}\ee since both $\La$ and $m_1$ have a lower bound, $\Lambda = 3\ \text{TeV}$ and $m_1= m_e$, implying \be |n|\simeq  0.7\times10^6.\label{nmad1}\ee This coincides with the $n$ value in Fig. \ref{fig3} at which the red line intersects either $\La=3$ TeV in the upper panel or $N_{\mathrm{sig}}/100=1$ in the lower panel.
\item Also from the figure, the model can generally maintain the excess in XENON1T with a stronger $B-L$ breaking scale and/or larger dark matter mass, but this requires the minimum value of $n$ to be correspondingly enhanced, which is indeed bigger than that in (\ref{nmad1}) by factors $(m_1/m_e)(\La/3\mathrm{TeV})^2$. For instance, we obtain \be |n| \simeq 1.94\times 10^6,\ 6.86\times 10^6,\ 1.9\times 10^7,\label{etad2}\ee according to the last three pairs of $(m_1,\La)$ in (\ref{etad1}), which can be seen in Fig. \ref{fig3} lower panel when the relevant curves intersect the dashed line. 

The nature of a large $U(1)$ charge and the behavior of its coupling are explained in Appendix \ref{u1chargecoupling}. From the remark 1 and 2, such a minimum value of $n$ would translate to an upper bound for $g_{B-L}$ and $m_{Z'}$, as shown below. 
\item Obviously, the model with $n=0$ fails to account for the excess, unless the kinetic mixing and $g_{B-L}$ fine-tuning to unnaturally small values is imposed, which is not interpreted in this work.     
\een

Applying the criteria in Appendix \ref{u1chargecoupling} to $U(1)_{B-L}$, we approximate $b_{B-L}\simeq -(5/3)(1/3+2n)^2$, thus \be \mu\pa g'_{B-L}/\pa\mu\simeq (5/48\pi^2)g'^3_{B-L},\ee which yields that $g'_{B-L}\equiv |1/3+2n|g_{B-L}$ slides similarly to the usual one. The solution is \be \al'^{-1}=\al'^{-1}_G-(5/6\pi)\ln(\mu/\mu_G),\ee where $\al'\equiv g'^2_{B-L}/4\pi$ and a subscript $_G$ referring to that of the GUT (the Planck regime may be chosen, but not necessary). Further, one demands a perturbative limit for the $U(1)_{B-L}$ gauge interaction at the GUT scale, i.e. $\al'_G\simeq 1$ at $\mu_G\sim 10^{16}$ GeV, which also necessarily prevents a proton decay since it occurs only via a GUT breakdown. This leads to $\al'\simeq 0.08$ at the scale of interest, $\mu\sim 1$~MeV.\footnote{Generally, the physical processes relevant to the gauge interactions of $F_2$ and $\Phi$ are determined by $\al'$, thus are perturbative, since $\al'$ decreases below 1 when the energy scale decreases below the GUT scale.} Thus, we get \be g'_{B-L}=|1/3+2n|g_{B-L}\simeq 1.\label{dntnd0}\ee 

According to the values of $n$ in (\ref{nmad1}) and (\ref{etad2}), the relation (\ref{dntnd0}) implies upper bounds for \bea g_{B-L} \simeq \fr{0.5}{n} &\simeq&  7.1\times 10^{-7},\ 2.6\times 10^{-7},\crn
&& 7.2\times 10^{-8},\ 2.6\times 10^{-8},\label{gadt1d}\eea respectively. With the aid of (\ref{dtlt3r}) and (\ref{dntnd0}), the $Z'$ mass is rewritten as \be m_{Z'} \simeq  3.62\times 10^{3}(g_{B-L}/m_1)^{1/2} \mathrm{MeV}^{3/2}.\label{ddt231}\ee Corresponding to the above values of $(g_{B-L},m_1)$, the $Z'$ mass is bounded, respectively, by  
\be m_{Z'} \simeq 4.3,\ 2.6,\ 0.43,\ 0.26\ \mathrm{MeV}.\label{zpadt2d}\ee Notice that if $(m_1,\La)$ are higher than those in (\ref{etad1}), the $Z'$ coupling and mass get more restricted. 

Since the $Z'$ mass is radically low, the constraints from low energy experiments apply. Because of the conditions for $m_{Z'}$ in (\ref{f2densityb2}) as well as $m_2 < m_0$, the $Z'$ boson cannot decay invisibly to any dark field, $\mathrm{Br}(Z'\rightarrow F^c_2 F_2) = \mathrm{Br}(Z'\rightarrow \Phi^* \Phi)=0$, and obviously $\mathrm{Br}(Z'\rightarrow F^c_1 F_1) = 0$. Hence, $Z'$ may only decay to the neutrinos $Z'\rightarrow \nu^c_L\nu_L$ and charged leptons $Z'\rightarrow e^+e^-$ if $m_{Z'}>2m_e$, being the same with the usual $U(1)_{B-L}$ model, studied in \cite{Harnik:2012ni,Bilmis:2015lja,Kaneta:2016vkq,Ilten:2018crw,Bauer:2018onh,Lindner:2018kjo,Okada:2020cue}. That said, the $U(1)_{B-L}$ gauge coupling is strongly constrained by the beam dump experiments to be roundly $g_{B-L}\sim 10^{-8}$ for $m_{Z'}=1$--10 MeV (cf. the figures 5 and 13 of \cite{Ilten:2018crw} and \cite{Bauer:2018onh}, respectively). Such bound excludes the values of $g_{B-L}$ in (\ref{gadt1d}) that correspond to $m_{Z'}>2m_e\simeq 1$ MeV given in (\ref{zpadt2d}). However, the bound does not apply to $g_{B-L}$ that reduces $m_{Z'}$ below 1 MeV, since the detectors in such experiments are only sensitive to the signal of $Z'\rightarrow e^+e^-$ decay. Precisely, with the aid of (\ref{ddt231}), the surviving condition $m_{Z'} < 1$ MeV leads to
\be g_{B-L}<3.9\times 10^{-8}\left(m_1/m_e\right). \ee This regime excludes the processes (i) in App. \ref{f2density} that set the $F_2$ density at $Z'$ resonance $m_{Z'}=2m_2 > 1$ MeV, despite $\Omega_{F_2} h^2\sim (4m^2_2-m^2_{Z'})^2/(m^2_2 m^2_{Z'})$ tending to zero as desirable, where it yields $g_{B-L}> 4\times 10^{-8}$ in tension with the bound $g_{B-L}\sim 10^{-8}$.

\begin{figure}[h]
\includegraphics[scale=0.35]{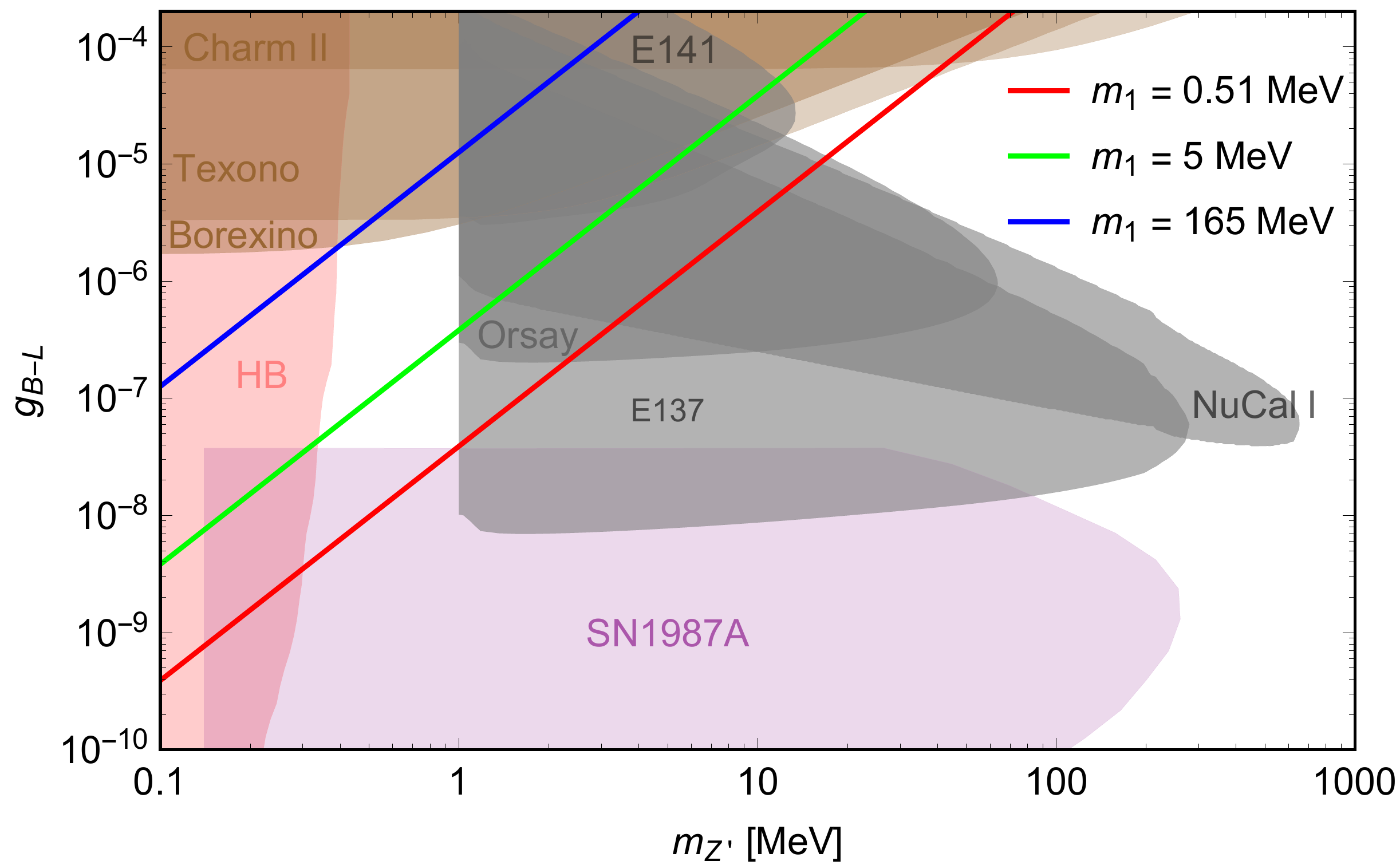}
\caption[]{\label{fig4} Theoretical prediction (\ref{ddt231}) for dark matter mass $m_1=0.51$, 5, and 165 MeV, where the relevant strongest experimental constraints (the color regions denote excluded parameter space according to each experiment) have been adapted from \cite{Ilten:2018crw} for the beam dump (E137, $\nu$-CAL I, Orsay, and E141), \cite{Bilmis:2015lja} for the neutrino-electron scattering (Borexino, Texono, and Charm~II with $\bar{\nu}_\mu$), \cite{Kaneta:2016vkq} for the HB stars (extracted from \cite{Redondo:2013lna}), and \cite{Harnik:2012ni} for the SN1987A (translated from \cite{Dent:2012mx}), where $m_1=165$~MeV was chosen so that the theoretical prediction intersects the HB and Borexino bounds.}
\end{figure}

Below $m_{Z'}=1$ MeV, the model is constrained by several experiments: The neutrino-electron scattering exchanged by $Z'$ fully studied in \cite{Bilmis:2015lja} using the measurements from Borexino \cite{Bellini:2011rx}, Texono \cite{Deniz:2009mu}, and Charm-II ($\bar{\nu}_\mu$) \cite{Vilain:1993kd,vilain:1994qy}; The energy loss, carried by $Z'$, of the horizontal branch (HB) stars \cite{Redondo:2013lna} and the supernova 1987A (SN1987A) \cite{Dent:2012mx,Kazanas:2014mca}. Including the beam dump experiments given by the darkcast package \cite{Ilten:2018crw} using the measurements from E137 \cite{Bjorken:1988as}, $\nu$-CAL I \cite{Blumlein:1990ay,Blumlein:1991xh}, Orsay \cite{Davier:1989wz}, and E141 \cite{Riordan:1987aw}, all the relevant experimental constraints and the model prediction from (\ref{ddt231}) are plotted in Figure~\ref{fig4}.
Hence, $m_1=165$ MeV is the upper limit of dark matter mass determined by the HB and Borexino bounds, while it is confidently that $m_1\simeq m_e\simeq 0.51$ MeV is the lower limit of dark matter mass at which the theory approximately coincides with the E137 and SN1987A bounds, that is \be 0.51\ \mathrm{MeV}< m_{1} <165\ \ \mathrm{MeV}.\ee Further, our model predicts \be 
0.34\ \mathrm{MeV} < m_{Z'}< 1 \ \mathrm{MeV},\ee limited by the HB and the beam dump insensitivity, and 
\be 3.8\times 10^{-8}<g_{B-L}<2.9\times 10^{-6},\ee bounded by the SN1987A and the Borexino, respectively. Notice that the region close to the top of Fig. \ref{fig4}, that is bounded by the red line $m_1=0.51$ MeV and allowed by the Texono and Orsay experiments, is excluded by the XENON1T, since $m_{Z'}\leq 4.3$ MeV.   

{\it Conclusion.} We have discovered a seminal result of the seesaw mechanism with gauged $B-L$ symmetry, alternative to the known leptogenesis, that it manifestly solves the long-standing issue of structured dark matter. The seesaw scale presents a nontrivial physical vacuum preserving two residual gauge symmetries, relating to the usual matter parity $P_M=(-1)^{3(B-L)+2s}$ and a new $Z_3$ quotient $[p^2]=[w^{3(B-L)}]$. This yields a two-component dark matter scenario naturally addressing the recent XENON1T excess. The cold dark matter $F_1$ has $B-L=0$, while the boosted dark matter $F_2$ has $B-L$ deviating from $1/3$ by seven order (roundly), and their masses $m_{1,2}$ obey $\mathrm{Max}(m_e,m_{Z'})<m_1\approx m_2<165$ MeV. The $Z'$ mass and coupling predicted are $0.34\ \mathrm{MeV}<m_{Z'}<1\ \mathrm{MeV}$ and $ 3.8\times 10^{-8}< g_{B-L}<2.9\times 10^{-6}$, respectively. If the XENON1T experiment is relaxed, a scheme of two-component dark matter beyond the weak scale is warranted, presented in standard model gauge extensions containing a $B-L$ charge, such as the left-right symmetry and $SO(10)$. 

{\it Acknowledgments.} This research is funded by Vietnam National Foundation for Science and Technology Development (NAFOSTED) under grant number 103.01-2019.353.

\appendix

\section{\label{appdat} High energy collider constraint}

Contribution of $Z'$ to the process $e^+ e^-\rightarrow f\bar{f}$, where $f$ is an ordinary fermion, proceeds through the $Z'$ exchange either by $s$-channel diagram for $f\neq e$ or by $s,t,u$-channel diagrams for $f=e$. 

If $m_{Z'}<\sqrt{s}=209$ GeV as in this model, the $Z'$ boson is resonantly produced at the LEPII and the agreement between the LEPII and the standard model indicates that the $Z'$-fermion couplings are $g_{B-L} \lesssim 10^{-2}$, given that $Z'$ does not decay to the dark fields $F_2,\Phi$ \cite{Appelquist:2002mw}. However, when $m_{Z'}>2m_{F_2,\Phi}$, the decay to the dark fields dominates the $Z'$ width and then the cross-section $\sigma(e^+e^-\rightarrow f\bar{f})=\sigma(e^+e^-\rightarrow Z')\mathrm{Br}(Z'\rightarrow f\bar{f})$ is strongly suppressed, since $\mathrm{Br}(Z'\rightarrow f\bar{f})\sim 1/n^2\ll 1$. This case requiring $g_{B-L}/n\lesssim 10^{-2}$ is obviously satisfied for the $n$ value of interest and $g_{B-L}$ in perturbative regime. 

By contrast, if $m_{Z'}>\sqrt{s}=209$~GeV, the process $e^+ e^-\rightarrow f\bar{f}$ receives an off-shell contribution of $Z'$, best described by the effective interaction, \be \mathcal{L}_{\mathrm{eff}}\supset \fr{g^2_{B-L} (B-L)_{f}}{m^2_{Z'}-s}(\bar{e}\ga^\nu e)(\bar{f}\ga_\nu f).\ee Considering $f=\mu,\tau$, the LEPII has limited the $Z'$ mass over coupling to be $m_{Z'}/g_{B-L}\geq 6$ TeV, which translates to the breaking scale $\La\geq 3$ TeV \cite{Alcaraz:2006mx,Carena:2004xs}. 

Additionally, the LHC has searched for dilepton signals $pp\rightarrow l\bar{l}$ as mediated by $Z'$, yielding a mass limit roundly $m_{Z'}=4$ TeV for the $Z'$ coupling identical to $Z$, which converts to a bound for $\La\sim m_{Z'}/2g\sim 3$ TeV, similar to the LEPII \cite{Aaboud:2017buh}. Here $Z'$ does not decay to the dark matter. Otherwise, the cross-section is more suppressed, similar to the LEPII case. 

As a matter of fact, $Z'$ cannot decay to the dark fields in this model [cf. (\ref{f2densityb2})]. The production of mono-$X$ or two-$X$'s final state (signature) recoils against large missing momentum (or energy) carried by dark matter is insensitive to the LEPII and LHC detectors, because of the suppression $\mathrm{Br}(Z'\rightarrow \mathrm{DM}\ \mathrm{DM})=0$.  

In short, these experiments give a reference of the new physics scale, say $\La=3$ TeV, used in the body text. 
 
\section{\label{u1chargecoupling} Nature of $U(1)$ charge and running coupling}

In contrast to the non-Abelian charges (e.g., the color and the weak isospin) that are constrained by the non-Abelian nature of a Lie algebra, $[T_j,T_k]=i f_{jkl}T_l$ with $\mathrm{Tr}(T_j T_k)\sim \de_{jk}$, the Abelian charges including electric charge are completely arbitrary, often chosen to describe observed charge values, while do not explain them. 

Indeed, the Hamiltonian of a $U(1)_X$ gauge theory preserves a scaling symmetry, $g_X\rightarrow g'_X=g_X/\la$ and $X\rightarrow X'=\la X$, where $g'_X$ and $X'$ are the respective coupling and charge after the transformation. We can work in a basis that $g_X$ is small, while $X$ is large, which leaves the physics unchanged. 

For instance, the running of $g_X$ with renormalization scale $\mu$ satisfies the RG equation, \be \mu \pa g_X/\pa \mu = \beta(g_X)=-(g^3_X/16\pi^2)b_X,\label{u1ccb1}\ee where the beta function at 1-loop level is given by \be b_X=-\fr 2 3 \sum_L X^2_L-\fr 2 3 \sum_R X^2_R-\fr 1 3 \sum_S X^2_S\ee as summed over the left/right chiral fermion and scalar fields, respectively. The equation (\ref{u1ccb1}) conserves the above scaling symmetry as a result, implying that a theory with large $X$ behaves similarly to the usual ones.

A character of the $U(1)$ theory is that the function $b_X<0$ for every $X$-charge. Therefore, its gauge coupling $g_X$ decreases when the energy scale $\mu$ decreases. Given that the theory is definite at a GUT or Planck scale, i.e. $g_X X\sim 1$, it must properly work to be perturbative and predictive below such large scale.        

\section{\label{f2density} Condition of vanished $F_2$ density}

Last, but not least, it is verified that the $F_2$ relic density should be negligible. In the early universe, the dark field $F_2$ might completely annihilate to (i) the standard model particles such as $F^c_2 F_2\rightarrow l^+l^-,\nu^c_L\nu_L$ via the ($s$-channel) $Z'$ portal or to (ii) the new gauge boson $F^c_2 F_2\rightarrow Z'Z'$ via $t$- and $u$-channel diagrams mediated by just $F_2$, if kinetically allowed. 

If $m_2<m_{Z'}$, the annihilation proceeds through the first processes (i), yielding the cross-sections \be \langle \sigma v_{\mathrm{rel}}\rangle_{F^c_2F_2\rightarrow e^+e^-,\nu^c_L\nu_L} \sim \left\{\begin{array}{cc}
 \left(\fr{m_{Z'}}{2\ \mathrm{MeV}}\right)^4\ \mathrm{pb}, & \mathrm{if} \ m_2>\fr 1 2 m_{Z'}\\
 \left(\fr{m_2}{1\ \mathrm{MeV}}\right)^4\ \mathrm{pb}, & \mathrm{if} \ m_2<\fr 1 2 m_{Z'}
\end{array}\right.\nn \ee with the help of (\ref{dtlt3r}) as well as $m_1 \approx m_2$, where the resonance $m_2=\fr 1 2 m_{Z'}$ need not necessarily be considered since it sets a lowest $F_2$ relic density. Although one vertex (namely $\bar{F}_2 F_2 Z'$) is enhanced by the coupling strength $g'_{B-L}=|1/3+2n|g_{B-L}\simeq 1$, these contributions are only effective if Min$\{m_2,\fr 1 2 m_{Z'}\}>1$ MeV. In this case, since $m_1\approx m_2>$ 1 MeV, we obtain the $Z'$ mass bound to be $m_{Z'}<2.2$~MeV due to (\ref{ddt231}), (\ref{dntnd0}), and (\ref{dtlt3r}) for $\La\geq 3$ TeV. That said, the above possibilities give a finite contribution to dark matter, since $\langle \sigma v_{\mathrm{rel}}\rangle_{F^c_2F_2\rightarrow e^+e^-,\nu^c_L\nu_L} \sim 1$ pb, except for the resonance regime at which the $F_2$ density almost vanishes. 

By contrast, if $m_2>m_{Z'}$, the last process (ii) produces a large annihilation cross-section to be \be \langle \sigma v_{\mathrm{rel}}\rangle_{F^c_2 F_2\rightarrow Z'Z'}\simeq \fr{|M|^2}{16\pi s}\gg 1\ \mathrm{pb},\ee because the amplitude $M(F^c_2 F_2\rightarrow Z'Z') \simeq 4g'^4_{B-L}$ that is induced in the order of unity is strongly enhanced by both the $\bar{F}_2 F_2 Z'$ vertices with the coupling strength $g'_{B-L}=|1/3+2n|g_{B-L}\simeq 1$ and that $s\simeq 4m^2_2 < \mathrm{GeV}^2$. 

Summarizing all the relevant cross-sections leads to $\Omega_{F_2}h^2\simeq 0.1\ \mathrm{pb}/\langle \sigma v_\mathrm{rel}\rangle_{F^c_2F_2\to \mathrm{all}}\ll 0.12$, provided that \be m_2>m_{Z'},\ \mathrm{or\ else}\ m_2=\fr 1 2 m_{Z'}.\label{f2densityb2}\ee 

\bibliographystyle{JHEP}
\bibliography{bibliography}

\providecommand{\href}[2]{#2}\begingroup\raggedright\begin{thebibliography}{10}

\bibitem{Tanabashi:2018oca}
{\scshape Particle Data Group} collaboration, \emph{{Review of Particle
  Physics}}, \href{https://doi.org/10.1103/PhysRevD.98.030001}{\emph{Phys.
  Rev.} {\bfseries D98} (2018) 030001}.

\bibitem{Minkowski:1977sc}
P.~Minkowski, \emph{{$\mu \to e\gamma$ at a Rate of One Out of $10^{9}$ Muon
  Decays?}}, \href{https://doi.org/10.1016/0370-2693(77)90435-X}{\emph{Phys.
  Lett.} {\bfseries 67B} (1977) 421}.

\bibitem{GellMann:1980vs}
M.~Gell-Mann, P.~Ramond and R.~Slansky, \emph{{Complex Spinors and Unified
  Theories}}, {\emph{Conf. Proc.} {\bfseries C790927} (1979) 315}
  [\href{https://arxiv.org/abs/1306.4669}{{\ttfamily 1306.4669}}].

\bibitem{Yanagida:1979as}
T.~Yanagida, \emph{{Horizontal symmetry and masses of neutrinos}}, {\emph{Conf.
  Proc.} {\bfseries C7902131} (1979) 95}.

\bibitem{Glashow:1979nm}
S.~L. Glashow, \emph{{The Future of Elementary Particle Physics}},
  \href{https://doi.org/10.1007/978-1-4684-7197-7_15}{\emph{NATO Sci. Ser. B}
  {\bfseries 61} (1980) 687}.

\bibitem{Mohapatra:1979ia}
R.~N. Mohapatra and G.~Senjanovic, \emph{{Neutrino Mass and Spontaneous Parity
  Violation}}, \href{https://doi.org/10.1103/PhysRevLett.44.912}{\emph{Phys.
  Rev. Lett.} {\bfseries 44} (1980) 912}.

\bibitem{Mohapatra:1980yp}
R.~N. Mohapatra and G.~Senjanovic, \emph{{Neutrino Masses and Mixings in Gauge
  Models with Spontaneous Parity Violation}},
  \href{https://doi.org/10.1103/PhysRevD.23.165}{\emph{Phys. Rev.} {\bfseries
  D23} (1981) 165}.

\bibitem{Lazarides:1980nt}
G.~Lazarides, Q.~Shafi and C.~Wetterich, \emph{{Proton Lifetime and Fermion
  Masses in an SO(10) Model}},
  \href{https://doi.org/10.1016/0550-3213(81)90354-0}{\emph{Nucl. Phys.}
  {\bfseries B181} (1981) 287}.

\bibitem{Schechter:1980gr}
J.~Schechter and J.~W.~F. Valle, \emph{{Neutrino Masses in SU(2) x U(1)
  Theories}}, \href{https://doi.org/10.1103/PhysRevD.22.2227}{\emph{Phys. Rev.}
  {\bfseries D22} (1980) 2227}.

\bibitem{Schechter:1981cv}
J.~Schechter and J.~W.~F. Valle, \emph{{Neutrino Decay and Spontaneous
  Violation of Lepton Number}},
  \href{https://doi.org/10.1103/PhysRevD.25.774}{\emph{Phys. Rev.} {\bfseries
  D25} (1982) 774}.

\bibitem{Davidson:1978pm}
A.~Davidson, \emph{{$B-L$ as the fourth color within an $SU(2)_L \times U(1)_R
  \times U(1)$ model}},
  \href{https://doi.org/10.1103/PhysRevD.20.776}{\emph{Phys. Rev.} {\bfseries
  D20} (1979) 776}.

\bibitem{Marshak:1979fm}
R.~E. Marshak and R.~N. Mohapatra, \emph{{Quark - Lepton Symmetry and B-L as
  the U(1) Generator of the Electroweak Symmetry Group}},
  \href{https://doi.org/10.1016/0370-2693(80)90436-0}{\emph{Phys. Lett.}
  {\bfseries 91B} (1980) 222}.

\bibitem{Mohapatra:1980qe}
R.~N. Mohapatra and R.~E. Marshak, \emph{{Local B-L Symmetry of Electroweak
  Interactions, Majorana Neutrinos and Neutron Oscillations}},
  \href{https://doi.org/10.1103/PhysRevLett.44.1644.2,
  10.1103/PhysRevLett.44.1316}{\emph{Phys. Rev. Lett.} {\bfseries 44} (1980)
  1316}.

\bibitem{Aprile:2020tmw}
{\scshape XENON} collaboration, \emph{{Excess electronic recoil events in
  XENON1T}}, \href{https://doi.org/10.1103/PhysRevD.102.072004}{\emph{Phys.
  Rev. D} {\bfseries 102} (2020) 072004}
  [\href{https://arxiv.org/abs/2006.09721}{{\ttfamily 2006.09721}}].

\bibitem{Krauss:1988zc}
L.~M. Krauss and F.~Wilczek, \emph{{Discrete Gauge Symmetry in Continuum
  Theories}}, \href{https://doi.org/10.1103/PhysRevLett.62.1221}{\emph{Phys.
  Rev. Lett.} {\bfseries 62} (1989) 1221}.

\bibitem{Martin:1992mq}
S.~P. Martin, \emph{{Some simple criteria for gauged R-parity}},
  \href{https://doi.org/10.1103/PhysRevD.46.R2769}{\emph{Phys. Rev. D}
  {\bfseries 46} (1992) 2769}
  [\href{https://arxiv.org/abs/hep-ph/9207218}{{\ttfamily hep-ph/9207218}}].

\bibitem{Batell:2010bp}
B.~Batell, \emph{{Dark Discrete Gauge Symmetries}},
  \href{https://doi.org/10.1103/PhysRevD.83.035006}{\emph{Phys. Rev. D}
  {\bfseries 83} (2011) 035006}
  [\href{https://arxiv.org/abs/1007.0045}{{\ttfamily 1007.0045}}].

\bibitem{Ma:2015xla}
E.~Ma, \emph{{Derivation of Dark Matter Parity from Lepton Parity}},
  \href{https://doi.org/10.1103/PhysRevLett.115.011801}{\emph{Phys. Rev. Lett.}
  {\bfseries 115} (2015) 011801}
  [\href{https://arxiv.org/abs/1502.02200}{{\ttfamily 1502.02200}}].

\bibitem{Ma:2015mjd}
E.~Ma, N.~Pollard, R.~Srivastava and M.~Zakeri, \emph{{Gauge $B-L$ Model with
  Residual $Z_3$ Symmetry}},
  \href{https://doi.org/10.1016/j.physletb.2015.09.010}{\emph{Phys. Lett. B}
  {\bfseries 750} (2015) 135}
  [\href{https://arxiv.org/abs/1507.03943}{{\ttfamily 1507.03943}}].

\bibitem{Hirsch:2017col}
M.~Hirsch, R.~Srivastava and J.~W.~F. Valle, \emph{{Can one ever prove that
  neutrinos are Dirac particles?}},
  \href{https://doi.org/10.1016/j.physletb.2018.03.073}{\emph{Phys. Lett. B}
  {\bfseries 781} (2018) 302}
  [\href{https://arxiv.org/abs/1711.06181}{{\ttfamily 1711.06181}}].

\bibitem{Bonilla:2018ynb}
C.~Bonilla, S.~Centelles-Chuliá, R.~Cepedello, E.~Peinado and R.~Srivastava,
  \emph{{Dark matter stability and Dirac neutrinos using only Standard Model
  symmetries}}, \href{https://doi.org/10.1103/PhysRevD.101.033011}{\emph{Phys.
  Rev. D} {\bfseries 101} (2020) 033011}
  [\href{https://arxiv.org/abs/1812.01599}{{\ttfamily 1812.01599}}].

\bibitem{Heeck:2013rpa}
J.~Heeck and W.~Rodejohann, \emph{{Neutrinoless Quadruple Beta Decay}},
  \href{https://doi.org/10.1209/0295-5075/103/32001}{\emph{EPL} {\bfseries 103}
  (2013) 32001} [\href{https://arxiv.org/abs/1306.0580}{{\ttfamily
  1306.0580}}].

\bibitem{Cai:2018nob}
C.~Cai, Z.~Kang, H.-H. Zhang and Y.-P. Zeng, \emph{{Minimal dark matter in
  $SU(2)_L x U(1)_Y x U(1)_{B-L}$}},
  \href{https://doi.org/10.1016/j.physletb.2018.08.014}{\emph{Phys. Lett. B}
  {\bfseries 784} (2018) 385}
  [\href{https://arxiv.org/abs/1801.05594}{{\ttfamily 1801.05594}}].

\bibitem{Nanda:2019nqy}
D.~Nanda and D.~Borah, \emph{{Connecting Light Dirac Neutrinos to a
  Multi-component Dark Matter Scenario in Gauged $B-L$ Model}},
  \href{https://doi.org/10.1140/epjc/s10052-020-8122-4}{\emph{Eur. Phys. J. C}
  {\bfseries 80} (2020) 557}
  [\href{https://arxiv.org/abs/1911.04703}{{\ttfamily 1911.04703}}].

\bibitem{Fukugita:1986hr}
M.~Fukugita and T.~Yanagida, \emph{{Baryogenesis Without Grand Unification}},
  \href{https://doi.org/10.1016/0370-2693(86)91126-3}{\emph{Phys. Lett.}
  {\bfseries B174} (1986) 45}.

\bibitem{Takahashi:2020bpq}
F.~Takahashi, M.~Yamada and W.~Yin, \emph{{XENON1T Excess from Anomaly-Free
  Axionlike Dark Matter and Its Implications for Stellar Cooling Anomaly}},
  \href{https://doi.org/10.1103/PhysRevLett.125.161801}{\emph{Phys. Rev. Lett.}
  {\bfseries 125} (2020) 161801}
  [\href{https://arxiv.org/abs/2006.10035}{{\ttfamily 2006.10035}}].

\bibitem{Kannike:2020agf}
K.~Kannike, M.~Raidal, H.~Veerm\"ae, A.~Strumia and D.~Teresi, \emph{{Dark
  Matter and the XENON1T electron recoil excess}},
  \href{https://doi.org/10.1103/PhysRevD.102.095002}{\emph{Phys. Rev. D}
  {\bfseries 102} (2020) 095002}
  [\href{https://arxiv.org/abs/2006.10735}{{\ttfamily 2006.10735}}].

\bibitem{Choi:2020udy}
G.~Choi, M.~Suzuki and T.~T. Yanagida, \emph{{XENON1T Anomaly and its
  Implication for Decaying Warm Dark Matter}},
  \href{https://arxiv.org/abs/2006.12348}{{\ttfamily 2006.12348}}.

\bibitem{Buch:2020mrg}
J.~Buch, M.~A. Buen-Abad, J.~Fan and J.~S.~C. Leung, \emph{{Galactic Origin of
  Relativistic Bosons and XENON1T Excess}},
  \href{https://arxiv.org/abs/2006.12488}{{\ttfamily 2006.12488}}.

\bibitem{Chen:2020gcl}
Y.~Chen, J.~Shu, X.~Xue, G.~Yuan and Q.~Yuan, \emph{{Sun Heated MeV-scale Dark
  Matter and the XENON1T Electron Recoil Excess}},
  \href{https://arxiv.org/abs/2006.12447}{{\ttfamily 2006.12447}}.

\bibitem{Bell:2020bes}
N.~F. Bell, J.~B. Dent, B.~Dutta, S.~Ghosh, J.~Kumar and J.~L. Newstead,
  \emph{{Explaining the XENON1T excess with Luminous Dark Matter}},
  \href{https://doi.org/10.1103/PhysRevLett.125.161803}{\emph{Phys. Rev. Lett.}
  {\bfseries 125} (2020) 161803}
  [\href{https://arxiv.org/abs/2006.12461}{{\ttfamily 2006.12461}}].

\bibitem{Du:2020ybt}
M.~Du, J.~Liang, Z.~Liu, V.~Q. Tran and Y.~Xue, \emph{{On-shell mediator dark
  matter models and the Xenon1T excess}},
  \href{https://doi.org/10.1088/1674-1137/abc244}{\emph{Chin. Phys. C}
  {\bfseries 45} (2021) 013114}
  [\href{https://arxiv.org/abs/2006.11949}{{\ttfamily 2006.11949}}].

\bibitem{Su:2020zny}
L.~Su, W.~Wang, L.~Wu, J.~M. Yang and B.~Zhu, \emph{{Atmospheric Dark Matter
  and Xenon1T Excess}},
  \href{https://doi.org/10.1103/PhysRevD.102.115028}{\emph{Phys. Rev. D}
  {\bfseries 102} (2020) 115028}
  [\href{https://arxiv.org/abs/2006.11837}{{\ttfamily 2006.11837}}].

\bibitem{Harigaya:2020ckz}
K.~Harigaya, Y.~Nakai and M.~Suzuki, \emph{{Inelastic Dark Matter Electron
  Scattering and the XENON1T Excess}},
  \href{https://doi.org/10.1016/j.physletb.2020.135729}{\emph{Phys. Lett. B}
  {\bfseries 809} (2020) 135729}
  [\href{https://arxiv.org/abs/2006.11938}{{\ttfamily 2006.11938}}].

\bibitem{Fornal:2020npv}
B.~Fornal, P.~Sandick, J.~Shu, M.~Su and Y.~Zhao, \emph{{Boosted Dark Matter
  Interpretation of the XENON1T Excess}},
  \href{https://doi.org/10.1103/PhysRevLett.125.161804}{\emph{Phys. Rev. Lett.}
  {\bfseries 125} (2020) 161804}
  [\href{https://arxiv.org/abs/2006.11264}{{\ttfamily 2006.11264}}].

\bibitem{Alonso-Alvarez:2020cdv}
G.~Alonso-\'Alvarez, F.~Ertas, J.~Jaeckel, F.~Kahlhoefer and L.~J. Thormaehlen,
  \emph{{Hidden Photon Dark Matter in the Light of XENON1T and Stellar
  Cooling}}, \href{https://doi.org/10.1088/1475-7516/2020/11/029}{\emph{JCAP}
  {\bfseries 11} (2020) 029}
  [\href{https://arxiv.org/abs/2006.11243}{{\ttfamily 2006.11243}}].

\bibitem{Jho:2020sku}
Y.~Jho, J.-C. Park, S.~C. Park and P.-Y. Tseng, \emph{{Leptonic New Force and
  Cosmic-ray Boosted Dark Matter for the XENON1T Excess}},
  \href{https://doi.org/10.1016/j.physletb.2020.135863}{\emph{Phys. Lett. B}
  {\bfseries 811} (2020) 135863}
  [\href{https://arxiv.org/abs/2006.13910}{{\ttfamily 2006.13910}}].

\bibitem{Baryakhtar:2020rwy}
M.~Baryakhtar, A.~Berlin, H.~Liu and N.~Weiner, \emph{{Electromagnetic Signals
  of Inelastic Dark Matter Scattering}},
  \href{https://arxiv.org/abs/2006.13918}{{\ttfamily 2006.13918}}.

\bibitem{Bloch:2020uzh}
I.~M. Bloch, A.~Caputo, R.~Essig, D.~Redigolo, M.~Sholapurkar and T.~Volansky,
  \emph{{Exploring New Physics with O(keV) Electron Recoils in Direct Detection
  Experiments}},  \href{https://arxiv.org/abs/2006.14521}{{\ttfamily
  2006.14521}}.

\bibitem{Paz:2020pbc}
G.~Paz, A.~A. Petrov, M.~Tammaro and J.~Zupan, \emph{{Shining dark matter in
  Xenon1T}},  \href{https://arxiv.org/abs/2006.12462}{{\ttfamily 2006.12462}}.

\bibitem{Cao:2020bwd}
Q.-H. Cao, R.~Ding and Q.-F. Xiang, \emph{{Exploring for sub-MeV Boosted Dark
  Matter from Xenon Electron Direct Detection}},
  \href{https://arxiv.org/abs/2006.12767}{{\ttfamily 2006.12767}}.

\bibitem{Lee:2020wmh}
H.~M. Lee, \emph{{Exothermic dark matter for XENON1T excess}},
  \href{https://doi.org/10.1007/JHEP01(2021)019}{\emph{JHEP} {\bfseries 01}
  (2021) 019} [\href{https://arxiv.org/abs/2006.13183}{{\ttfamily
  2006.13183}}].

\bibitem{Nakayama:2020ikz}
K.~Nakayama and Y.~Tang, \emph{{Gravitational Production of Hidden Photon Dark
  Matter in Light of the XENON1T Excess}},
  \href{https://doi.org/10.1016/j.physletb.2020.135977}{\emph{Phys. Lett. B}
  {\bfseries 811} (2020) 135977}
  [\href{https://arxiv.org/abs/2006.13159}{{\ttfamily 2006.13159}}].

\bibitem{Primulando:2020rdk}
R.~Primulando, J.~Julio and P.~Uttayarat, \emph{{Collider Constraints on a Dark
  Matter Interpretation of the XENON1T Excess}},
  \href{https://doi.org/10.1140/epjc/s10052-020-08652-x}{\emph{Eur. Phys. J. C}
  {\bfseries 80} (2020) 1084}
  [\href{https://arxiv.org/abs/2006.13161}{{\ttfamily 2006.13161}}].

\bibitem{Gelmini:2020xir}
G.~B. Gelmini, V.~Takhistov and E.~Vitagliano, \emph{{Scalar direct detection:
  In-medium effects}},
  \href{https://doi.org/10.1016/j.physletb.2020.135779}{\emph{Phys. Lett. B}
  {\bfseries 809} (2020) 135779}
  [\href{https://arxiv.org/abs/2006.13909}{{\ttfamily 2006.13909}}].

\bibitem{Bramante:2020zos}
J.~Bramante and N.~Song, \emph{{Electric But Not Eclectic: Thermal Relic Dark
  Matter for the XENON1T Excess}},
  \href{https://doi.org/10.1103/PhysRevLett.125.161805}{\emph{Phys. Rev. Lett.}
  {\bfseries 125} (2020) 161805}
  [\href{https://arxiv.org/abs/2006.14089}{{\ttfamily 2006.14089}}].

\bibitem{Zu:2020idx}
L.~Zu, G.-W. Yuan, L.~Feng and Y.-Z. Fan, \emph{{Mirror Dark Matter and
  Electronic Recoil Events in XENON1T}},
  \href{https://arxiv.org/abs/2006.14577}{{\ttfamily 2006.14577}}.

\bibitem{Baek:2020owl}
S.~Baek, J.~Kim and P.~Ko, \emph{{XENON1T excess in local $Z_2$ DM models with
  light dark sector}},
  \href{https://doi.org/10.1016/j.physletb.2020.135848}{\emph{Phys. Lett. B}
  {\bfseries 810} (2020) 135848}
  [\href{https://arxiv.org/abs/2006.16876}{{\ttfamily 2006.16876}}].

\bibitem{Alhazmi:2020fju}
H.~Alhazmi, D.~Kim, K.~Kong, G.~Mohlabeng, J.-C. Park and S.~Shin,
  \emph{{Implications of the XENON1T Excess on the Dark Matter
  Interpretation}},  \href{https://arxiv.org/abs/2006.16252}{{\ttfamily
  2006.16252}}.

\bibitem{Chao:2020yro}
W.~Chao, Y.~Gao and M.~j. Jin, \emph{{Pseudo-Dirac Dark Matter in XENON1T}},
  \href{https://arxiv.org/abs/2006.16145}{{\ttfamily 2006.16145}}.

\bibitem{DelleRose:2020pbh}
L.~Delle~Rose, G.~H\"utsi, C.~Marzo and L.~Marzola, \emph{{Impact of
  loop-induced processes on the boosted dark matter interpretation of the
  XENON1T excess}},
  \href{https://doi.org/10.1088/1475-7516/2021/02/031}{\emph{JCAP} {\bfseries
  02} (2021) 031} [\href{https://arxiv.org/abs/2006.16078}{{\ttfamily
  2006.16078}}].

\bibitem{Ko:2020gdg}
P.~Ko and Y.~Tang, \emph{{Semi-annihilating $Z_3$ dark matter for XENON1T
  excess}}, \href{https://doi.org/10.1016/j.physletb.2021.136181}{\emph{Phys.
  Lett. B} {\bfseries 815} (2021) 136181}
  [\href{https://arxiv.org/abs/2006.15822}{{\ttfamily 2006.15822}}].

\bibitem{An:2020tcg}
H.~An and D.~Yang, \emph{{Direct detection of freeze-in inelastic dark
  matter}},  \href{https://arxiv.org/abs/2006.15672}{{\ttfamily 2006.15672}}.

\bibitem{Okada:2020evk}
N.~Okada, S.~Okada, D.~Raut and Q.~Shafi, \emph{{Dark matter $Z^\prime$ and
  XENON1T excess from $U(1)_X$ extended standard model}},
  \href{https://doi.org/10.1016/j.physletb.2020.135785}{\emph{Phys. Lett. B}
  {\bfseries 810} (2020) 135785}
  [\href{https://arxiv.org/abs/2007.02898}{{\ttfamily 2007.02898}}].

\bibitem{Choudhury:2020xui}
D.~Choudhury, S.~Maharana, D.~Sachdeva and V.~Sahdev, \emph{{Dark matter, muon
  anomalous magnetic moment, and the XENON1T excess}},
  \href{https://doi.org/10.1103/PhysRevD.103.015006}{\emph{Phys. Rev. D}
  {\bfseries 103} (2021) 015006}
  [\href{https://arxiv.org/abs/2007.08205}{{\ttfamily 2007.08205}}].

\bibitem{Arcadi:2020zni}
G.~Arcadi, A.~Bally, F.~Goertz, K.~Tame-Narvaez, V.~Tenorth and S.~Vogl,
  \emph{{EFT interpretation of XENON1T electron recoil excess: Neutrinos and
  dark matter}}, \href{https://doi.org/10.1103/PhysRevD.103.023024}{\emph{Phys.
  Rev. D} {\bfseries 103} (2021) 023024}
  [\href{https://arxiv.org/abs/2007.08500}{{\ttfamily 2007.08500}}].

\bibitem{He:2020wjs}
H.-J. He, Y.-C. Wang and J.~Zheng, \emph{{EFT Approach of Inelastic Dark Matter
  for Xenon Electron Recoil Detection}},
  \href{https://doi.org/10.1088/1475-7516/2021/01/042}{\emph{JCAP} {\bfseries
  01} (2021) 042} [\href{https://arxiv.org/abs/2007.04963}{{\ttfamily
  2007.04963}}].

\bibitem{Dey:2020sai}
U.~K. Dey, T.~N. Maity and T.~S. Ray, \emph{{Prospects of Migdal Effect in the
  Explanation of XENON1T Electron Recoil Excess}},
  \href{https://doi.org/10.1016/j.physletb.2020.135900}{\emph{Phys. Lett. B}
  {\bfseries 811} (2020) 135900}
  [\href{https://arxiv.org/abs/2006.12529}{{\ttfamily 2006.12529}}].

\bibitem{Smirnov:2020zwf}
J.~Smirnov and J.~F. Beacom, \emph{{New Freezeout Mechanism for Strongly
  Interacting Dark Matter}},
  \href{https://doi.org/10.1103/PhysRevLett.125.131301}{\emph{Phys. Rev. Lett.}
  {\bfseries 125} (2020) 131301}
  [\href{https://arxiv.org/abs/2002.04038}{{\ttfamily 2002.04038}}].

\bibitem{Choi:2020kch}
G.~Choi, T.~T. Yanagida and N.~Yokozaki, \emph{{Feebly interacting $U
  (1)_{B-L}$ gauge boson warm dark matter and XENON1T anomaly}},
  \href{https://doi.org/10.1016/j.physletb.2020.135836}{\emph{Phys. Lett. B}
  {\bfseries 810} (2020) 135836}
  [\href{https://arxiv.org/abs/2007.04278}{{\ttfamily 2007.04278}}].

\bibitem{Miranda:2020kwy}
O.~G. Miranda, D.~K. Papoulias, M.~T\'ortola and J.~W.~F. Valle, \emph{{XENON1T
  signal from transition neutrino magnetic moments}},
  \href{https://doi.org/10.1016/j.physletb.2020.135685}{\emph{Phys. Lett. B}
  {\bfseries 808} (2020) 135685}
  [\href{https://arxiv.org/abs/2007.01765}{{\ttfamily 2007.01765}}].

\bibitem{Lindner:2020kko}
M.~Lindner, Y.~Mambrini, T.~B. de~Melo and F.~S. Queiroz, \emph{{XENON1T
  anomaly: A light Z' from a Two Higgs Doublet Model}},
  \href{https://doi.org/10.1016/j.physletb.2020.135972}{\emph{Phys. Lett. B}
  {\bfseries 811} (2020) 135972}
  [\href{https://arxiv.org/abs/2006.14590}{{\ttfamily 2006.14590}}].

\bibitem{AristizabalSierra:2020edu}
D.~Aristizabal~Sierra, V.~De~Romeri, L.~Flores and D.~Papoulias, \emph{{Light
  vector mediators facing XENON1T data}},
  \href{https://arxiv.org/abs/2006.12457}{{\ttfamily 2006.12457}}.

\bibitem{Boehm:2020ltd}
C.~Boehm, D.~G. Cerdeno, M.~Fairbairn, P.~A. Machado and A.~C. Vincent,
  \emph{{Light new physics in XENON1T}},
  \href{https://arxiv.org/abs/2006.11250}{{\ttfamily 2006.11250}}.

\bibitem{Khan:2020vaf}
A.~N. Khan, \emph{{Can Nonstandard Neutrino Interactions explain the XENON1T
  spectral excess?}},
  \href{https://doi.org/10.1016/j.physletb.2020.135782}{\emph{Phys. Lett. B}
  {\bfseries 809} (2020) 135782}
  [\href{https://arxiv.org/abs/2006.12887}{{\ttfamily 2006.12887}}].

\bibitem{Bally:2020yid}
A.~Bally, S.~Jana and A.~Trautner, \emph{{Neutrino self-interactions and
  XENON1T electron recoil excess}},
  \href{https://doi.org/10.1103/PhysRevLett.125.161802}{\emph{Phys. Rev. Lett.}
  {\bfseries 125} (2020) 161802}
  [\href{https://arxiv.org/abs/2006.11919}{{\ttfamily 2006.11919}}].

\bibitem{Sabti:2019mhn}
N.~Sabti, J.~Alvey, M.~Escudero, M.~Fairbairn and D.~Blas, \emph{{Refined
  Bounds on MeV-scale Thermal Dark Sectors from BBN and the CMB}},
  \href{https://doi.org/10.1088/1475-7516/2020/01/004}{\emph{JCAP} {\bfseries
  01} (2020) 004} [\href{https://arxiv.org/abs/1910.01649}{{\ttfamily
  1910.01649}}].

\bibitem{Joglekar:2019vzy}
A.~Joglekar, N.~Raj, P.~Tanedo and H.-B. Yu, \emph{{Relativistic capture of
  dark matter by electrons in neutron stars}},
  \href{https://arxiv.org/abs/1911.13293}{{\ttfamily 1911.13293}}.

\bibitem{Agashe:2014yua}
K.~Agashe, Y.~Cui, L.~Necib and J.~Thaler, \emph{{(In)direct Detection of
  Boosted Dark Matter}},
  \href{https://doi.org/10.1088/1475-7516/2014/10/062}{\emph{JCAP} {\bfseries
  10} (2014) 062} [\href{https://arxiv.org/abs/1405.7370}{{\ttfamily
  1405.7370}}].

\bibitem{Harnik:2012ni}
R.~Harnik, J.~Kopp and P.~A.~N. Machado, \emph{{Exploring nu Signals in Dark
  Matter Detectors}},
  \href{https://doi.org/10.1088/1475-7516/2012/07/026}{\emph{JCAP} {\bfseries
  1207} (2012) 026} [\href{https://arxiv.org/abs/1202.6073}{{\ttfamily
  1202.6073}}].

\bibitem{Bilmis:2015lja}
S.~Bilmis, I.~Turan, T.~M. Aliev, M.~Deniz, L.~Singh and H.~T. Wong,
  \emph{{Constraints on Dark Photon from Neutrino-Electron Scattering
  Experiments}}, \href{https://doi.org/10.1103/PhysRevD.92.033009}{\emph{Phys.
  Rev. D} {\bfseries 92} (2015) 033009}
  [\href{https://arxiv.org/abs/1502.07763}{{\ttfamily 1502.07763}}].

\bibitem{Kaneta:2016vkq}
K.~Kaneta, Z.~Kang and H.-S. Lee, \emph{{Right-handed neutrino dark matter
  under the $B-L$ gauge interaction}},
  \href{https://doi.org/10.1007/JHEP02(2017)031}{\emph{JHEP} {\bfseries 02}
  (2017) 031} [\href{https://arxiv.org/abs/1606.09317}{{\ttfamily
  1606.09317}}].

\bibitem{Ilten:2018crw}
P.~Ilten, Y.~Soreq, M.~Williams and W.~Xue, \emph{{Serendipity in dark photon
  searches}}, \href{https://doi.org/10.1007/JHEP06(2018)004}{\emph{JHEP}
  {\bfseries 06} (2018) 004}
  [\href{https://arxiv.org/abs/1801.04847}{{\ttfamily 1801.04847}}].

\bibitem{Bauer:2018onh}
M.~Bauer, P.~Foldenauer and J.~Jaeckel, \emph{{Hunting All the Hidden
  Photons}}, \href{https://doi.org/10.1007/JHEP07(2018)094}{\emph{JHEP}
  {\bfseries 18} (2020) 094}
  [\href{https://arxiv.org/abs/1803.05466}{{\ttfamily 1803.05466}}].

\bibitem{Lindner:2018kjo}
M.~Lindner, F.~S. Queiroz, W.~Rodejohann and X.-J. Xu, \emph{{Neutrino-electron
  scattering: general constraints on $Z'$ and dark photon models}},
  \href{https://doi.org/10.1007/JHEP05(2018)098}{\emph{JHEP} {\bfseries 05}
  (2018) 098} [\href{https://arxiv.org/abs/1803.00060}{{\ttfamily
  1803.00060}}].

\bibitem{Okada:2020cue}
N.~Okada, S.~Okada and Q.~Shafi, \emph{{Light $Z'$ and dark matter from
  U(1)$_X$ gauge symmetry}},
  \href{https://doi.org/10.1016/j.physletb.2020.135845}{\emph{Phys. Lett. B}
  {\bfseries 810} (2020) 135845}
  [\href{https://arxiv.org/abs/2003.02667}{{\ttfamily 2003.02667}}].

\bibitem{Bellini:2011rx}
G.~Bellini et~al., \emph{{Precision measurement of the 7Be solar neutrino
  interaction rate in Borexino}},
  \href{https://doi.org/10.1103/PhysRevLett.107.141302}{\emph{Phys. Rev. Lett.}
  {\bfseries 107} (2011) 141302}
  [\href{https://arxiv.org/abs/1104.1816}{{\ttfamily 1104.1816}}].

\bibitem{Deniz:2009mu}
{\scshape TEXONO collaboration} collaboration, \emph{{Measurement of
  Neutrino-Electron Scattering Cross-Section with a CsI(Tl) Scintillating
  Crystal Array at the Kuo-Sheng Nuclear Power Reactor}}, {\emph{Phys. Rev.}
  {\bfseries D81} (2010) 072001}
  [\href{https://arxiv.org/abs/0911.1597}{{\ttfamily 0911.1597}}].

\bibitem{Vilain:1993kd}
{\scshape CHARM-II} collaboration, \emph{{Measurement of differential
  cross-sections for muon-neutrino electron scattering}},
  \href{https://doi.org/10.1016/0370-2693(93)90408-A}{\emph{Phys. Lett.}
  {\bfseries B302} (1993) 351}.

\bibitem{vilain:1994qy}
{\scshape CHARM-II} collaboration, \emph{Precision measurement of electroweak
  parameters from the scattering of muon-neutrinos on electrons}, {\emph{Phys.
  Lett.} {\bfseries B335} (1994) 246}.

\bibitem{Redondo:2013lna}
J.~Redondo and G.~Raffelt, \emph{{Solar constraints on hidden photons
  re-visited}},
  \href{https://doi.org/10.1088/1475-7516/2013/08/034}{\emph{JCAP} {\bfseries
  08} (2013) 034} [\href{https://arxiv.org/abs/1305.2920}{{\ttfamily
  1305.2920}}].

\bibitem{Dent:2012mx}
J.~B. Dent, F.~Ferrer and L.~M. Krauss, \emph{{Constraints on Light Hidden
  Sector Gauge Bosons from Supernova Cooling}},
  \href{https://arxiv.org/abs/1201.2683}{{\ttfamily 1201.2683}}.

\bibitem{Kazanas:2014mca}
D.~Kazanas, R.~N. Mohapatra, S.~Nussinov, V.~L. Teplitz and Y.~Zhang,
  \emph{{Supernova Bounds on the Dark Photon Using its Electromagnetic Decay}},
  \href{https://doi.org/10.1016/j.nuclphysb.2014.11.009}{\emph{Nucl. Phys. B}
  {\bfseries 890} (2014) 17} [\href{https://arxiv.org/abs/1410.0221}{{\ttfamily
  1410.0221}}].

\bibitem{Bjorken:1988as}
J.~D. Bjorken, S.~Ecklund, W.~R. Nelson, A.~Abashian, C.~Church, B.~Lu et~al.,
  \emph{{Search for Neutral Metastable Penetrating Particles Produced in the
  SLAC Beam Dump}}, \href{https://doi.org/10.1103/PhysRevD.38.3375}{\emph{Phys.
  Rev. D} {\bfseries 38} (1988) 3375}.

\bibitem{Blumlein:1990ay}
J.~Blumlein et~al., \emph{{Limits on neutral light scalar and pseudoscalar
  particles in a proton beam dump experiment}},
  \href{https://doi.org/10.1007/BF01548556}{\emph{Z. Phys. C} {\bfseries 51}
  (1991) 341}.

\bibitem{Blumlein:1991xh}
J.~Blumlein et~al., \emph{{Limits on the mass of light (pseudo)scalar particles
  from Bethe-Heitler e+ e- and mu+ mu- pair production in a proton - iron beam
  dump experiment}},
  \href{https://doi.org/10.1142/S0217751X9200171X}{\emph{Int. J. Mod. Phys. A}
  {\bfseries 7} (1992) 3835}.

\bibitem{Davier:1989wz}
M.~Davier and H.~Nguyen~Ngoc, \emph{{An Unambiguous Search for a Light Higgs
  Boson}}, \href{https://doi.org/10.1016/0370-2693(89)90174-3}{\emph{Phys.
  Lett. B} {\bfseries 229} (1989) 150}.

\bibitem{Riordan:1987aw}
E.~M. Riordan et~al., \emph{{A Search for Short Lived Axions in an Electron
  Beam Dump Experiment}},
  \href{https://doi.org/10.1103/PhysRevLett.59.755}{\emph{Phys. Rev. Lett.}
  {\bfseries 59} (1987) 755}.

\bibitem{Appelquist:2002mw}
T.~Appelquist, B.~A. Dobrescu and A.~R. Hopper, \emph{{Nonexotic Neutral Gauge
  Bosons}}, \href{https://doi.org/10.1103/PhysRevD.68.035012}{\emph{Phys. Rev.
  D} {\bfseries 68} (2003) 035012}
  [\href{https://arxiv.org/abs/hep-ph/0212073}{{\ttfamily hep-ph/0212073}}].

\bibitem{Alcaraz:2006mx}
{\scshape ALEPH, DELPHI, L3, OPAL, LEP Electroweak Working Group}
  collaboration, \emph{{A Combination of preliminary electroweak measurements
  and constraints on the standard model}},
  \href{https://arxiv.org/abs/hep-ex/0612034}{{\ttfamily hep-ex/0612034}}.

\bibitem{Carena:2004xs}
M.~Carena, A.~Daleo, B.~A. Dobrescu and T.~M.~P. Tait, \emph{{$Z^\prime$ gauge
  bosons at the Tevatron}},
  \href{https://doi.org/10.1103/PhysRevD.70.093009}{\emph{Phys. Rev. D}
  {\bfseries 70} (2004) 093009}
  [\href{https://arxiv.org/abs/hep-ph/0408098}{{\ttfamily hep-ph/0408098}}].

\bibitem{Aaboud:2017buh}
{\scshape ATLAS} collaboration, \emph{{Search for new high-mass phenomena in
  the dilepton final state using 36 fb$^{-1}$ of proton-proton collision data
  at $ \sqrt{s}=13 $ TeV with the ATLAS detector}},
  \href{https://doi.org/10.1007/JHEP10(2017)182}{\emph{JHEP} {\bfseries 10}
  (2017) 182} [\href{https://arxiv.org/abs/1707.02424}{{\ttfamily
  1707.02424}}].

\end{thebibliography}\endgroup

\end{document}